\documentclass[useAMS,usenatbib]{mn2e}
\usepackage{graphicx,amsmath}
\usepackage{amssymb}
\usepackage{wrapfig}
\usepackage{esint}
\usepackage{multirow}

\usepackage[british,english]{babel}

\title[Thermal Comptonization Time Delays]{Energy Dependent Time Delays of kHz 
Oscillations 
due to Thermal Comptonization}

\author[Kumar and Misra]{Nagendra Kumar$^{1}$\thanks{E-mail:nagendrak@iucaa.ernet.in} and Ranjeev Misra$^{1}$\\
$^{1}$\textit{Inter-University Centre For Astronomy and Astrophysics, Post Bag4, Ganeshkind, 
Pune-411007, India }}

\begin{document}
%
\date{}

\pagerange{\pageref{firstpage}--\pageref{lastpage}} 

\maketitle

\label{firstpage}

\begin{abstract}
We study the energy dependent photon variability from a
thermal Comptonizing plasma that is oscillating at kHz frequencies.
In particular, we solve the linearised time dependent Kompaneets equation 
and consider the oscillatory perturbation to be either in the soft photon source or
in the heating rate of the plasma. For each case, we self consistently consider
the energy balance of the plasma and the soft photon source. The model
incorporates the possibility of a fraction of the Comptonized photons impinging
back into the soft photon source. We find that when the oscillation is
due to the soft photon source, the variation of the fractional root mean sqaure
(r.m.s)
is nearly constant with energy and the time-lags are hard. However, for the case
when the oscillation is due to  variation in the heating rate of the corona,
and  when a significant fraction of
the photons impinge back into the soft photon source, the r.m.s increases
with energy and the time lags are soft. As an example, we compare the results 
with
the $\sim 850$ Hz oscillation observed on March 3, 1996 for 4U 1608-52 and show 
that both the observed 
soft time-lags as well as the r.m.s versus energy can be well described  by
such a  model where the size of the Comptonizing plasma is $\sim 1$ km. Thus,
modelling of the time lags as due to Comptonization delays, 
 can provide tight constraints on
the size and geometry of the system. Detailed analysis would require well
constrained spectral parameters. 
\end{abstract}

\section{Introduction} 
X-ray binaries harbour compact objects and their 
X-ray luminosity  is
believed to be generated due to an  accretion disk 
formed by material accreted from a companion star. The X-ray emission
is known to vary on a wide range of timescales. 
It was  after the launch of the {\it Rossi 
X-ray Timing Explorer} (RXTE) satellite, that  millisecond variability
was detected in  several sources 
\citep[for reviews see for e.g.][]{vanderKlis2000,vanderKlis2006a,
Remillard-McClintock2006}. 
The X-ray variability is typically quantified in the frequency domain 
by computing
the  power density spectra. The power density spectrum sometimes shows peaked
features which are termed as quasi-periodic oscillations (QPO) 
In some neutron star low-mass X-ray binaries (LMXBs), there are 
QPOs observed in the frequency range of  400 to 1300 Hz, and these are 
termed as kilohertz QPOs. 
Often two simultaneous kHz QPOs are observed in a system, with 
the higher-frequency one being called  the `upper kHz QPO' while the
lower-frequency one
is called the `lower kHz QPO'. In general, 
the quality factor, $Q$ , which is the ratio of the QPO frequency to the
full width at half maximum, of the lower kHz QPO is significantly 
greater than the upper one \citep{Mendez2006}.    

There have been several models invoked to explain this twin kHz QPO
phenomena. A list of different models with their merits and demerits
has been presented by \cite{Lin-etal2011}. Here we briefly mention
some of them.  
In the sonic-point model 
\citep{Miller-Lamb-Psaltis1998,Lamb-Miller-Psaltis1998}
with modifications \citep{Lamb-Miller2003}, the  
frequency of the upper kHz QPO is interpreted as the  Keplerian one at the
sonic radius,  while  the frequency of lower kHz 
QPO is generated by the modulation of the upper QPO with the neutron star
spin frequency. 
In this interpretation, the frequency separation between the upper and lower 
kHz QPO should be a constant, but almost all sources show significant 
variation in the separation. For example, for the LMXB  4U 1608-52 a 
linear relation between the upper and lower kHz QPO has been reported
\citep{Mendez-etal1998b,Belloni-Mendez-Homan2005,Belloni-Mendez-Homan2007}.
The relativistic precession model \citep{Stella-Vietri1999} identifies 
the upper kHz QPO frequency as a Keplerian one  with the lower kHz QPO 
generated due to modulation of the Keplerian frequency with the relativistic 
periastron  precession frequency.
In the  two-oscillator model \citep{Osherovich-Titarchuk1999,
Titarchuk-Osherovich-Kuznestov1999},  the lower kHz QPO is due to a blob
moving in a Keplerian orbit, and the upper kHz QPO is generated by
the influence of the  Coriolis force on the blob when it enters the 
neutron star's rotating magnetosphere. Later,
\cite{Titarchuk2003} refined this model by connecting the QPO to a 
Rayleigh-Taylor gravity wave.
\citet{Fragile-Mathews-Wilson2001} have proposed that the 
accretion disk around a rapidly rotating compact object may be misaligned 
beyond a transition  radius due to the Bardeen-Petterson 
effect and this gives rise to the QPO phenomenon. 

There have also been attempts to understand these QPOs as global
oscillation in the disk due to hydrodynamic or magneto-hydrodynamic (MHD)
instabilities or waves. In a series of papers, 
\citet[e.g.][]{Kato2007,Kato2009} the QPO is explained to be 
generated by a resonant excitation caused by the  
non-linear interactions between the disk oscillation modes.  
The phenomena has been attributed  to Alfven wave oscillations 
\citep{Zhang2004} or to standing modes of MHD waves
\citep{Li-Zhang2005}. The Rossby wave instability in the inner disk
has also been invoked \citep{Lovelace-Turner-Romanova2009,
Lovelace-Romanova2007}. \citet{Shi-Li2009} have tried to explain the QPOs 
in terms of a resonant
coupling of MHD oscillation modes with the neutron star spin.
\cite{Bachetti-etal2010} have performed MHD simulations to understand 
the phenomena.   

Apart from theoretical consistences, the validity of these different
models can in principle be tested with observations. A model should
explain the observed relationships between the  twin kHz QPO frequencies and
any other lower frequency QPOs \citep[e.g.][]{Straaten-vanderKlis-Mendez2003,
Altamirano-etal2008}. It may also need to explain the strength
of the kHz QPOs  \citep[e.g.][]{Barret-Olive-Miller2005,
Mendez2006} or the condition for the appearance of two kHz QPOs
\citep[e.g.][]{Sanna-etal2012,Mendez2006,Lin-etal2012,Mendez-etal1998b}.
Unfortunately, there is no consensus that any of the models is the
most favourable one. This is primarily because none of the models have
a clear predictive and distinguishing signature which can be tested 
against these observations. 

It is also not clear how does a dynamical oscillation proposed by these 
different models, couples to the radiative process that produces the
X-ray emission. Relatively fewer studies have been undertaken to 
identify the radiative mechanism that is responsible for the kHz QPOs.
On the other hand, the time averaged spectra of LMXBs have been well
studied by several missions. The primary component has been indisputably
identified to be thermal Comptonization which typically dominates
the spectrum especially at energies $> 3$ keV. There is a secondary soft 
competent which 
could be a multi-coloured disk emission \citep[e.g.][]{Mit84,Mit89,Dis02,Agr09} 
or a black body from the boundary layer between the disk and the star
\citep[e.g.][]{Whi86,Dis01}. The spectra show significant and systematic
long term variation which is characterised by shapes (``Z'' and ``atoll'')
in the colour-colour and colour-intensity plots. These variations are
primarily due to the main thermal component. The temperature of the
Comptonizing medium ranges from $\sim 2$ keV (``soft'' state) to
$ \sim 15$ keV (``hard'' state) while corresponding electron
scattering optical depth ranges from $\sim 10$ to $ \sim 2$
\citep[e.g.][]{Rai11}.

The occurrence of kHz QPO has been reported to be related to the
spectral shape or more specifically to the position of the source in
its colour-colour diagram \citep{Straaten-vanderKlis-Mendez2003,
Altamirano-etal2008, Wijnands-etal1997,Lin-etal2012}.  The detection
of the QPOs in the long term light curves of sources have also
been examined \citep{Yu-etal1997,Mendez-etal1998b,Belloni-etal2007}.
These studies seem to suggest that the QPOs are preferentially detected
in intermediate spectral states which lie between the extreme soft
and hard states. On the other hand, statistically analysis undertaken
by \citet{Mis04} suggests that the fraction of segments having a kHz
QPO is near unity at low states and decreases to zero at high states
with nearly constant amplitude. In any case, irrespective of the details
mentioned above, the kHz QPO is most probably  associated with the primary
thermal Comptonization component which typically dominates the spectra.

The fractional root mean square (r.m.s) amplitude of the kHz QPOs have been 
measured 
to be an increasing function of energy  
\citep{Berger-etal1996,Mendez-vanderKlis-Ford2001,Zhang-etal1996,
Wijnands-etal1997} 
with  indications of a decrease at $ > 20 $ keV \citep{Muk12}. This
again implies that the phenomenon should be associated with the
high energy thermal Comptonization component and not the soft black body one.
Remarkable evidence was obtained when the time lag between different energy
bins for the lower kHz QPO  was measured to be of the order of
$\sim 30$ micro-seconds \citep{Vaughan-etal1997}. For 
thermal Comptonization, higher energy photons scatter more than 
the low energy ones and hence one expects time-lags between them. For a 
corona of size of few  kms and optical depths of order unity, the
expected time difference is of the order of $\sim 50$ micro-seconds. It
was realised that the variation of the r.m.s and the time-lag as a function
of energy is a powerful diagnostic which can distinguish which parameter
(e.g. the plasma temperature, the soft photon input) drives the oscillation as well as
 constrain the size and geometry of the system \citep{Lee-Miller1998}. 
For the LMXB  4U 1608-52, the
time lag was found to be  ``soft'' i.e. the soft photons are delayed
compared to the hard ones \citep[erratum of][]{Vaughan-etal1997}. Such soft lags 
were confirmed also for 4U 1636-53 \citep{Kaaret-etal1999}. This is
in apparent contradiction to the thermal Comptonization model, where
one would expect hard lags. However, \cite{Lee-Misra-Taam2001} showed
that if a fraction of the Comptonized photons impinge back in the
soft photon source, then the system would show soft lags with an
energy dependence similar to what has been observed.

Earlier, time lags for kHz QPOs were reported for a single 
RXTE observation of 4U 1608-52 \citep{Vaughan-etal1997}
and 4U 1636-53 \citep{Kaaret-etal1999}. Recently \cite{deAvellar-etal2013}
and \cite{Barret2013} has analysed a number of observations of 
both sources and have confirmed that the time lag is soft and is
of the order of tens of micro-secs for the lower kHz QPO. The time
lag remains nearly constant till about $\sim 800$ Hz and then decreases
as the frequency increases.  \cite{deAvellar-etal2013} have studied for
the first time the time-lag for the upper kHz QPO and find them to be
hard lags and inconsistent with those seen for the lower kHz QPO. 
Thus, these results ascertain that time-lags in kHz QPOs is a rich
phenomenon which can provide important insight into the nature of these
systems.

Apart from delays due to electron scatterings in a thermal plasma,
the time-lags could be due to other reasons. Longer time-lags ($> 10$ msec) 
observed in the low frequency continuum \citep{Nowak-etal1999} are 
attributed to propagation of 
stochastic variability in a disk \citep{Lyu97,Mis00,Kot01}
or for a low frequency QPO \citep[e.g.][]{Cui99}, time-lag may be due 
to non-linear dependence of spectral parameters with each other \citep{Mis13}.
The micro-second time-lags observed for kHz QPOs are unlikely to be
due to any of the  above but could be due to reverberation of the X-rays
from a reflector which is  few kms away from the source \citep{Barret2013}.
However, it should be emphasised that since the primary spectral component
is due thermal Comptonization, time lag due to multiple electron scattering
must occur and should be of the order of tens of micro-seconds.
Thus, while reverberation may contribute to the observed time-lags, the
Comptonization lag is an important effect, which cannot be neglected for
kHz QPOs.

In this work, we study using the linearised time dependent Kompaneets
equation, the response of the thermal plasma to variations in 
the input photon flux and the plasma heating rate. For the latter,
we also consider the possibility that a fraction of photons impinge back
into the input photon source. The analysis predicts the shape of the r.m.s
as a function of energy as well as the expected time-lags as a function
of energy. We apply the method to a test case and show that such 
analysis can provide constraints on the geometry and size of the system.

In the next section we develop the mathematical formulation which is
followed by \S 3 where the generic results are presented. In \S 4, 
as a test case,
we apply the method to data from 4U 1608-52 
to show the utility of such an analysis. In \S 5 we summarise and discuss
the work.

\section{Variability of Thermal Comptonized photons }

Thermal Comptonization is a process in which low energy ``seed''  photons are
Comptonized by a hot thermal electron gas. While the geometry of the 
Comptonizing system in 
an X-ray binary may be complex, here we assume a simple one,
where a spherical
seed photon source of radius $a$ is surrounded by a Comptonizing medium 
of width $L$ and temperature $T_e$. The input source is considered to be
a black body with temperature $T_b$ and hence the rate of input photons per 
unit volume 
inside the medium is given as 
\begin{align}
\dot{n}_{s\gamma}= \left[\frac{3  a^2}{[(a+L)^3-a^3]}\right] 
\left( \frac{2\pi}{h^3c^2} \frac{E^2}{(\exp\left[{\frac{E}{kT_b}}\right]-1)}
\right)
\end{align}
The Thompson collision time scale is 
$t_c = 1/(c n_e \sigma_T)$ and the optical depth of the medium
is $\tau = L n_e \sigma_T$, where $\sigma_T$ is
Thompson cross-section and $n_e$ is the electron density.
The evolution of the photon density, $n_\gamma$, inside the Comptonized medium
in the non-relativistic limit (i.e. kT$_e$ $\ll$ m$_e c^2$) and for low
photon energies (E $\ll$ m$_e c^2$) is governed by the Kompaneets equation 
\citep{Kom57}
where the induced scattering term is neglected:
\begin{align}
t_c\frac{dn_\gamma}{dt} =& \frac{1}{m_ec^2}\frac{d}{dE}
\left[-4kT_eEn_\gamma + E^2n_\gamma + kT_e \frac{d}{dE}(E^2n_\gamma)
\right]  \nonumber \\ &
+ t_c\dot{n}_{s\gamma} - t_c\dot{n}_{esc}
\end{align}
It should be noted that Eqn (2) describes the evolution of
the photon density, $n_\gamma$, while typically the Kompaneets equation
is often written in terms of the photon occupation number
$n \propto n_\gamma/E^2$. In this work it is more intuitive to
work directly with the photon density instead of the
occupation number.
Here 
$\dot{n}_{esc}$ is the rate of escape of the photon density which is taken to be 
$\dot{n}_{esc} \simeq$ $\dfrac{n_\gamma}{(\tau^2+\tau)t_c}$. In steady
state,  the equilibrium photon density $n_{\gamma o}$ can be computed by
setting $\frac{dn_\gamma}{dt} = 0$ and  $\dot{n}_{esc o}$ represents the
emergent spectrum. The spectral shape is determined by the corresponding steady
state values $T_{eo}$, $\tau_{o}$ and $T_{bo}$. 

The electron temperature of the corona is set by the balance
of Compton cooling and some external heating. In general the time
evolution of the temperature is governed by
\begin{equation}
\frac{3}{2}k\frac{\partial T_e}{\partial t} 
= \dot{H}_{Ex} - {\langle \Delta \dot E \rangle}
\label{Heat}
\end{equation}
where $\dot{H}_{Ex}$ is the external heating rate per electron and
the Compton cooling rate per electron is given by
\begin{equation}
{\langle \Delta \dot E \rangle} = \int_{E_{min}}^{E^{max}}(4kT_e-E)
\frac{E}{m_ec^2}\ n_{\gamma} \ \sigma_T \ c\ dE
\end{equation}
In steady state, when the electron temperature is $T_{eo}$, 
$\dot{\langle \Delta E \rangle_o} = \dot{H}_{Exo}$.

We take into account the possibility that a fraction, $\eta$, of the
Comptonized photons impinge back onto to the seed photon source,
thereby increasing the temperature of the seed photon source which
can be estimated by 
\begin{equation} 
4 \pi a^2 \sigma T_{b}^4 =4 \pi a^2 \sigma T_{b}^{'4}  + \eta V_c \int 
\frac{n_{\gamma }}{(\tau^2+\tau)t_c} E \ dE 
\label{feedback}
\end{equation}
where  
$V_c = (4/3) \pi {[(a+L)^3-a^3]}$ is the volume of the Comptonizing medium
and $T'_b$ is the input source temperature if $\eta = 0$ i.e.
$4 \pi a^2 \sigma (T'_{b})^4$ is the energy generation rate inside the
seed photon source. If the variability amplitude of a source is small,
then the time averaged spectrum will correspond to the steady state one
with parameters, $T_{eo}$, $\tau_o$ and $T_{bo}$.
For these values, $T'_{bo}$ can be estimated to be
\begin{equation}
T_{bo}^{'4} =  T_{bo}^4 - \frac{\eta {V_c}}{ 4 \pi a^2 \sigma} \int 
\frac{n_{\gamma_o }}{(\tau^2+\tau)t_c} E \ dE 
\end{equation}
and for the system there is a maximum allowed value of $\eta$, $\eta_{max}$ for
which $T'_{bo} = 0$.

We consider small amplitude oscillation of the medium electron
temperature and the source photon temperature over their average values
i.e. $T_e = T_{eo}(1+\Delta T_e \ e^{-i\omega t})$,
and $T_b = T_{bo} (1+ \Delta  T_b \ e^{-i\omega t})$. 
Here, $\omega$ is
the angular frequency of the oscillation and $\Delta T_e << 1$ and
$\Delta T_b << 1$ are in general complex quantities.
In the linear 
approximation, these will lead
to variation in the photon density as $n_\gamma = n_{\gamma o}
(1+ \Delta n_\gamma \ e^{-i\omega t})$. The oscillation amplitude of 
$n_\gamma$, $\Delta n_\gamma$ is related
to $\Delta T_e$ and $\Delta T_b$ by the linearised
Kompaneets equation, which is written in simplified form as 
\citep[e.g.][]{Lee-Miller1998,Lee-Misra-Taam2001}
\begin{align}
-&\frac{d^2\Delta n_\gamma}{dE^2}+\left(\frac{-1}{kT_{eo}}-
\frac{2}{n_{\gamma o}}\frac{dn_{\gamma o}}{dE}\right)
\frac{d\Delta n_{\gamma}}{dE} \nonumber \\ &
+\frac{m_ec^2t_c(\dot{n}_{s\gamma o}-i\omega n_{\gamma o})}
{E^2n_{\gamma o}kT_{eo}}\Delta n_\gamma =  
\left(\frac{-2}{E^2}+\frac{1}{n_{\gamma o}}
\frac{d^2n_{\gamma o}}{dE^2}\right)\Delta T_e \nonumber \\ &+
\frac{m_ec^2t_c\dot{n}_{s\gamma o}}{E^2n_{\gamma o}kT_{eo}}
\left(\frac{\frac{E}{kT_{bo}}}{1-\exp{\left(\frac{-E}{kT_{bo}}
\right)}}\right)\Delta T_b
\label{Komlin}
\end{align}
Here, $\dot{n}_{s\gamma o}$ is the time averaged input photon rate. Solving 
Eqn \ref{Komlin} one gets the complex variation $\Delta n_\gamma (E)$ as a
function of energy, which can be used to compute the fractional r.m.s with 
energy,
$|\Delta n_\gamma (E)|$, as well as phase lag between two energies $E_1$ and
$E_2$ which is the argument of 
$[\Delta n_\gamma(E_1)\Delta n^*_\gamma(E_2)]$.

We consider two possible drivers of the variability of the system (i)
variation in the seed photon temperature $T_b$ and (ii)  a
variation in the external heating rate of the corona $\dot{H}_{Ex}$
and analyse the induced
temporal variation for both cases.

\subsection{Variation of the seed photon temperature}

A variation in the seed photon temperature i.e. 
$T_b = T_{bo}(1+ \Delta  T_b \ e^{-i\omega t})$ could be due 
to variation in the energy generation rate
in the soft photon source. This will lead to a variation in the
photon number density $\Delta n_\gamma$ as governed by Eqn \ref{Komlin}.
The variation of the photon density would change the Compton cooling rate,
leading to a variation in medium temperature, $\Delta T_e$ which can
be estimated by linearising Eqn (\ref{Heat}) i.e.
\begin{align}
\frac{3}{2}kT_{eo}\Delta T_e (i\omega) &= \frac{\sigma_T c}{m_ec^2}
\left[\int 4kT_{eo}
(\Delta T_e \right. \nonumber \\ & \left. +\Delta n_{\gamma})
n_{\gamma o}\ EdE -\int E^2\Delta n_{\gamma}n_{\gamma o}dE\right]
\label{Heatlin}
\end{align}
where we have assumed that the external heating rate of the corona
is a constant at the steady state value,  $\dot{H}_{Ex} =\dot{H}_{Exo} $.

The two complex differential equations (\ref{Komlin}) and  (\ref{Heatlin}) have
to be solved simultaneously to obtain $\Delta T_e$  and 
$\Delta n_{\gamma}$ with
the latter being a function of energy. We use an iterative numerical scheme 
to solve them.
We start by having $\Delta T_e =0$ and obtain $\Delta n_{\gamma}$  from 
Eqn (\ref{Komlin}).
This is then used to get an updated value of $\Delta T_e$ by Eqn 
(\ref{Heatlin}). The 
procedure is iterated till $\Delta T_e$ converges. For all computations 
done here, we find
that the convergence is rapid and the converged value of $\Delta T_e$ is 
obtained within a few iterations.

\subsection{Variation of the coronal heating rate}

A variation of the coronal heating i.e. $\dot{H}_{Ex} = \dot{H}_{Exo} 
(1+ \Delta \dot{H}_{Ex} \ e^{-i\omega t})$.
would lead to a variation in the coronal temperature which is given by the 
linearised form of
Eqn (\ref{Heat}) i.e.
\begin{align}
\frac{3}{2}kT_{eo}\Delta T_e (-i\omega) &= \dot{H}_{Exo}
\Delta \dot{H}_{Ex}  -\frac{\sigma_T c}{m_ec^2}\left[\int 4kT_{eo}
(\Delta T_e \right. \nonumber \\ & \left. +\Delta n_{\gamma})
n_{\gamma o}\ EdE -\int E^2\Delta n_{\gamma}n_{\gamma o}dE\right]
\label{Heatlin2}
\end{align}
The induced variation in the temperature would naturally lead to variations
in the photon density. There is additional complication because the photon
density variation itself may lead to variations in the soft photon 
temperature, since
a fraction of them are assumed to impinge back on to the soft photon source.  
This effect
can be estimated by linearizing Eqn (\ref{feedback}),
\begin{equation}
4 \sigma (T_{bo})^4 \Delta T_b = \frac{\eta V_c}
{4 \pi a^2}
\int \frac{n_{\gamma o}}{(\tau^2+\tau)t_c}
\Delta n_{\gamma} \ E \ dE 
\label{feedbacklin}
\end{equation}

The three complex differential equations (\ref{Komlin}), (\ref{Heatlin2}) and 
(\ref{feedbacklin}) have
to be solved simultaneously to obtain $\Delta T_e$, $\Delta T_b$  
and $\Delta n_{\gamma}$. The
solutions can be readily obtained in an iterative manner which converges 
rapidly. 

\section{Results}

\begin{figure}\vspace{-1.7cm}
\begin{center}$
\begin{array}{cc}\vspace{-1.60cm}\hspace{-0.35cm}
\includegraphics[width=0.265\textwidth]{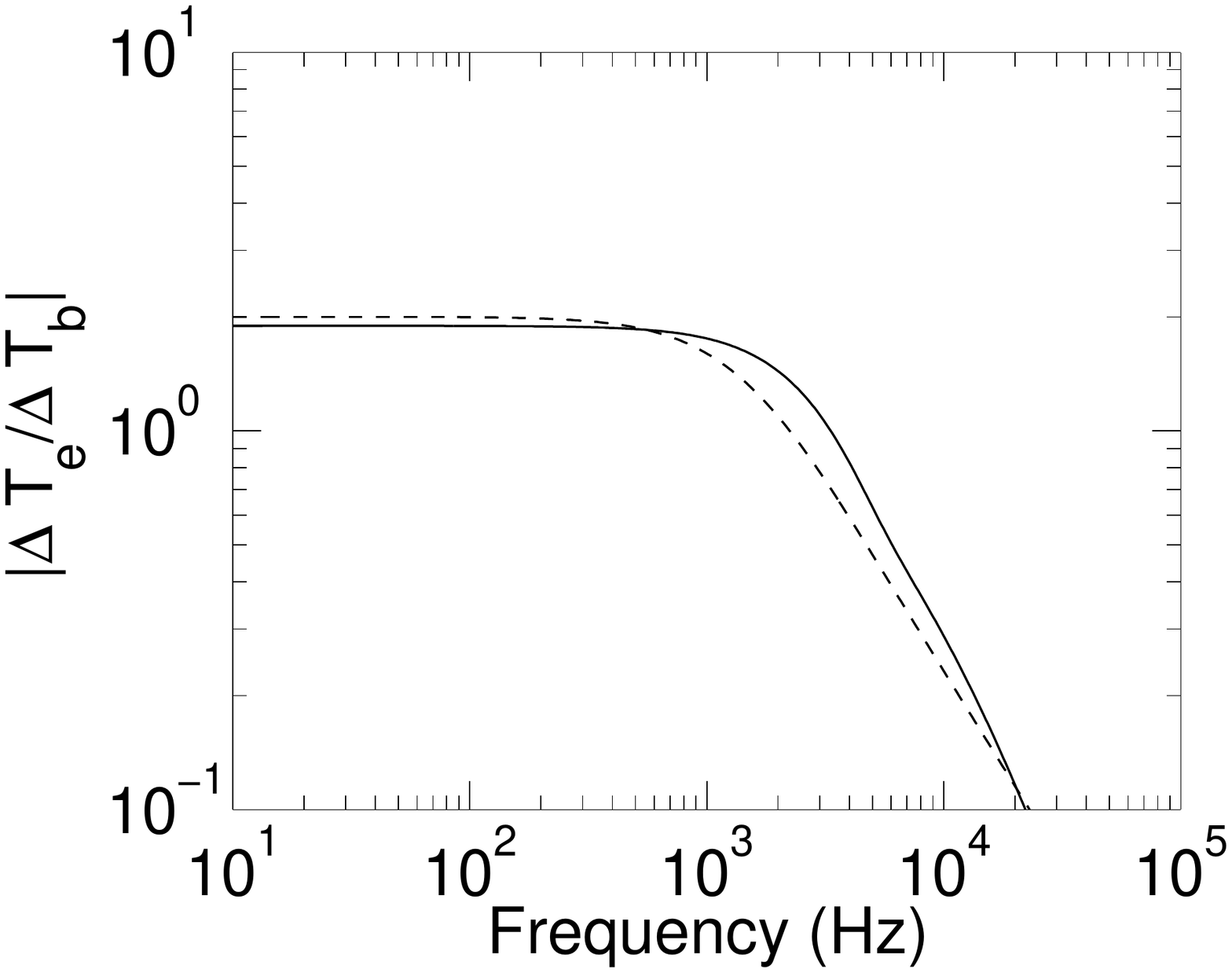}&\hspace{-0.6cm}
\includegraphics[width=0.265\textwidth]{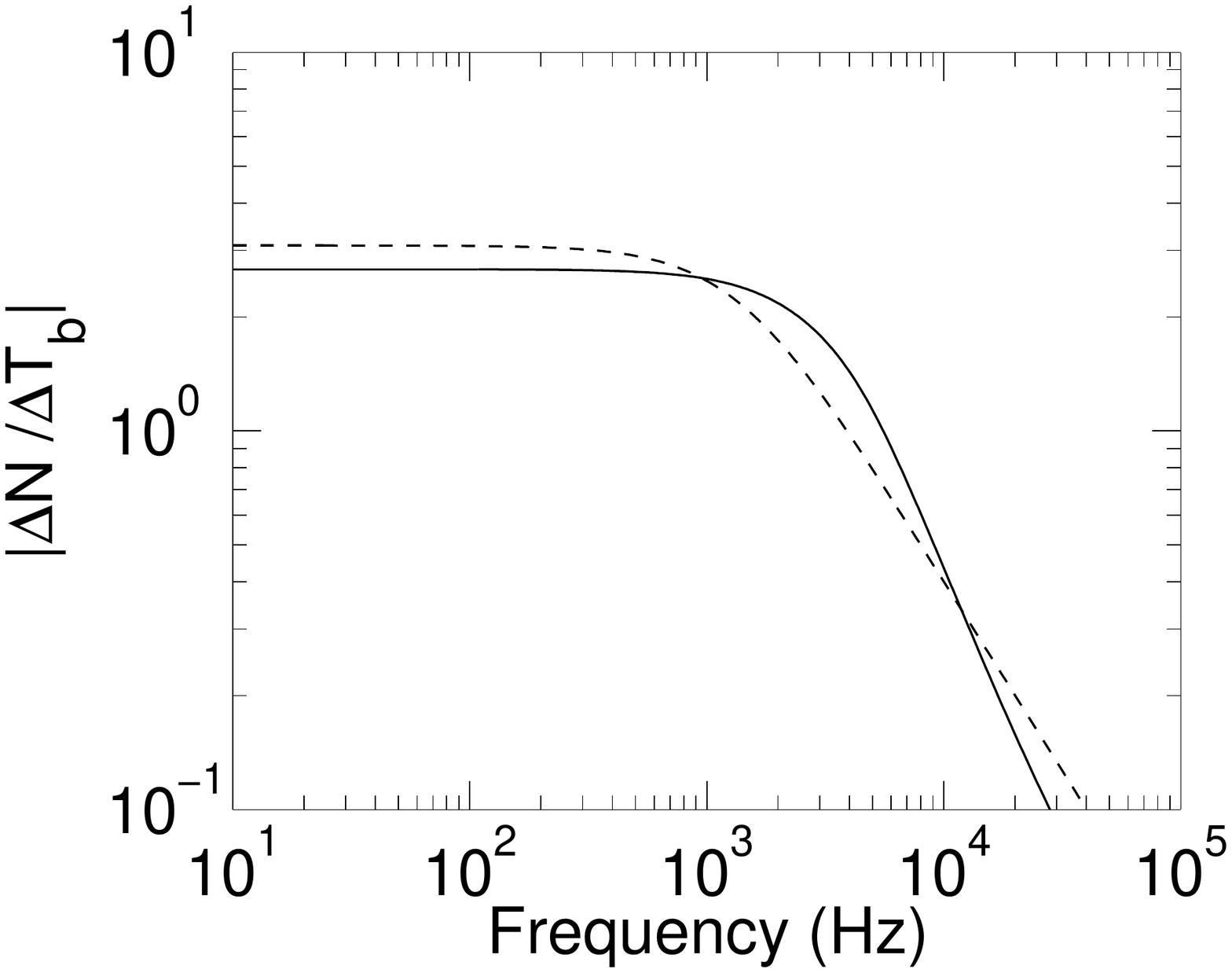}\\
\end{array}$
\end{center}
\caption{Variation of $| \Delta T_e/ \Delta T_b  |$ and the energy 
integrated normalised
photon density variation $ | \Delta N/ \Delta T_b  | $ as
a function of frequency. Here $\Delta N$ is the normalised integrated 
photon density variation in the energy band $2$ - $60$ keV
i.e. $\Delta N = \int n_\gamma \Delta n_\gamma dE/ \int n_\gamma dE$. 
The solid lines curves are for the ``hard  spectrum (Model A) and the
dashed lines curves are for the ``soft  spectrum (Model B).}
\label{seedtemp_freq}
\end{figure}

\begin{figure}\vspace{-1.7cm}
\begin{center}$
\begin{array}{cc}\vspace{-1.60cm}\hspace{-0.35cm}
\includegraphics[width=0.265\textwidth]{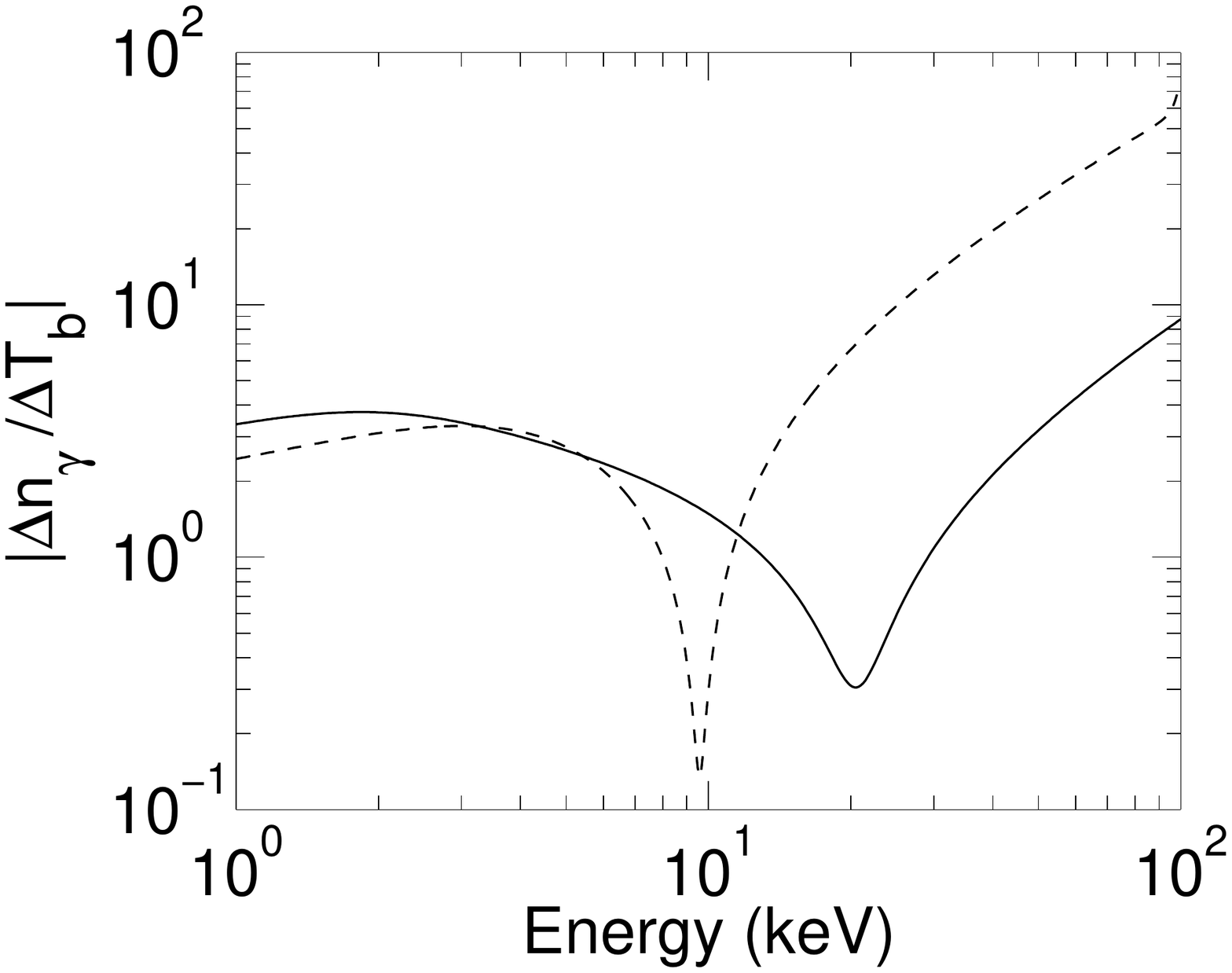}&\hspace{-0.6cm}
\includegraphics[width=0.265\textwidth]{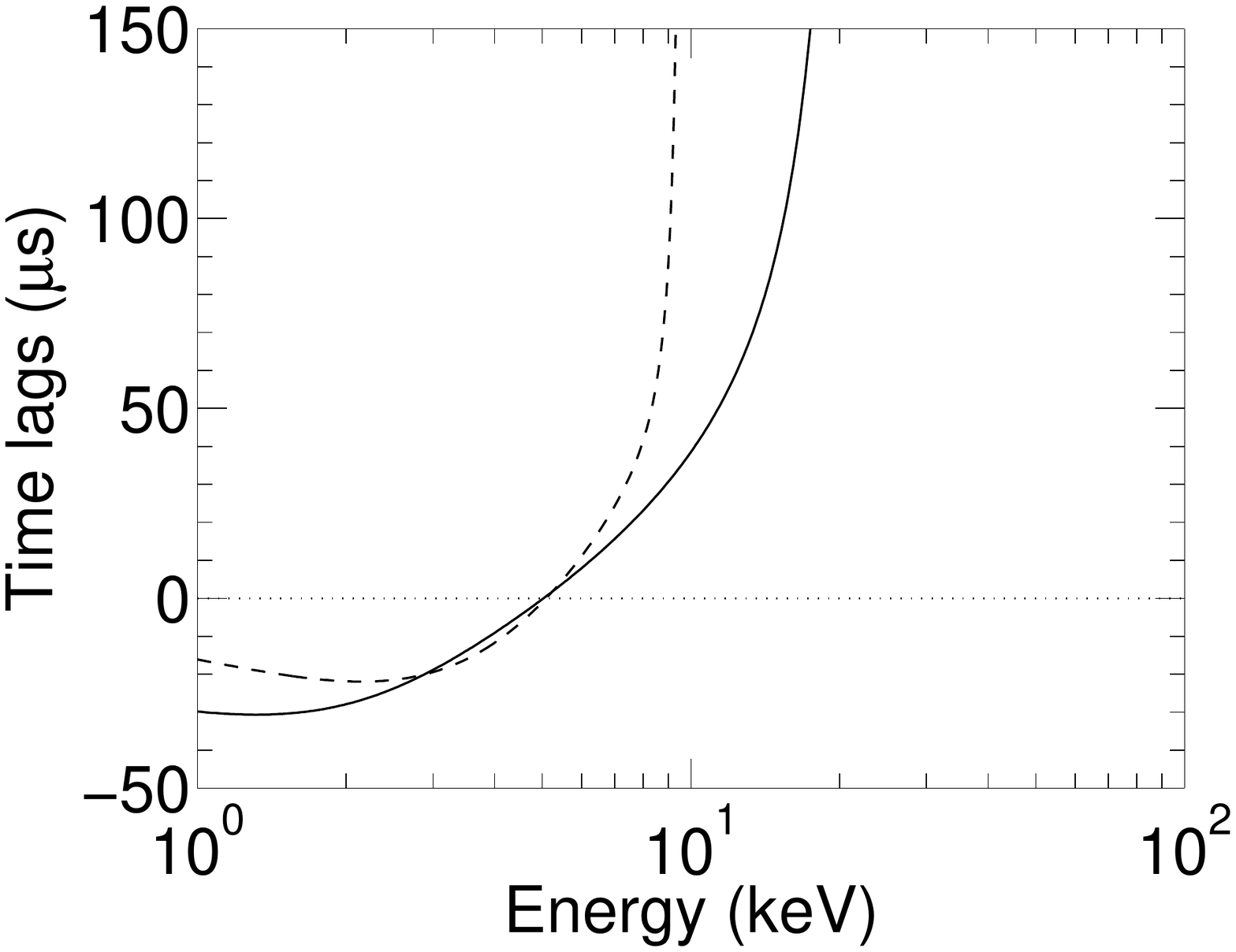}\\
\end{array}$
\end{center}
\caption{Variation of $| \Delta n_\gamma/ \Delta T_b  |$ 
and time-lag as a function of
photon energy. The solid lines curves are for the ``hard  spectrum (Model A) 
and the
dashed lines curves marked B are for the ``soft  spectrum (Model B).
The pivot point for both models is at high energies 
$E \sim 10$-$20$ keV and the time lags increase with energy 
which means that the hard photon lag the soft ones (i.e. hard lag).}
\label{seedtemp_Energy}
\end{figure}

To illustrate and to understand the energy dependent variability of the
Comptonized photons, we consider two different steady state spectra.
The first is a ``hard''  spectrum  (Model A) characterised by relatively 
high temperature and low optical depth:  $kT_e = 15$ keV,
$\tau^2 = 6$ and a seed photon temperature of $T_b = 0.3$ keV. The
second is a ``soft''  spectrum (Model B) with low temperature and high optical 
depth,
$kT_e = 2.5$ keV,
$\tau^2 = 36$ and a seed photon temperature of $T_b = 0.6$ keV. 
In practise,
these time averaged values can be obtained by fitting the average spectrum with
a steady state Comptonization model. Here we use these fiduciary parameter 
values
to bring out qualitative differences between any hard and soft average spectra.
For both models we choose the seed photon size $a = 10$ km and the corona size
to be $L = 5$ km and the frequency of oscillation to be $850$ Hz.

\begin{figure}\vspace{-1.7cm}
\begin{center}$
\begin{array}{cc}\vspace{-2.8cm}\hspace{-0.35cm}
\includegraphics[width=0.265\textwidth]{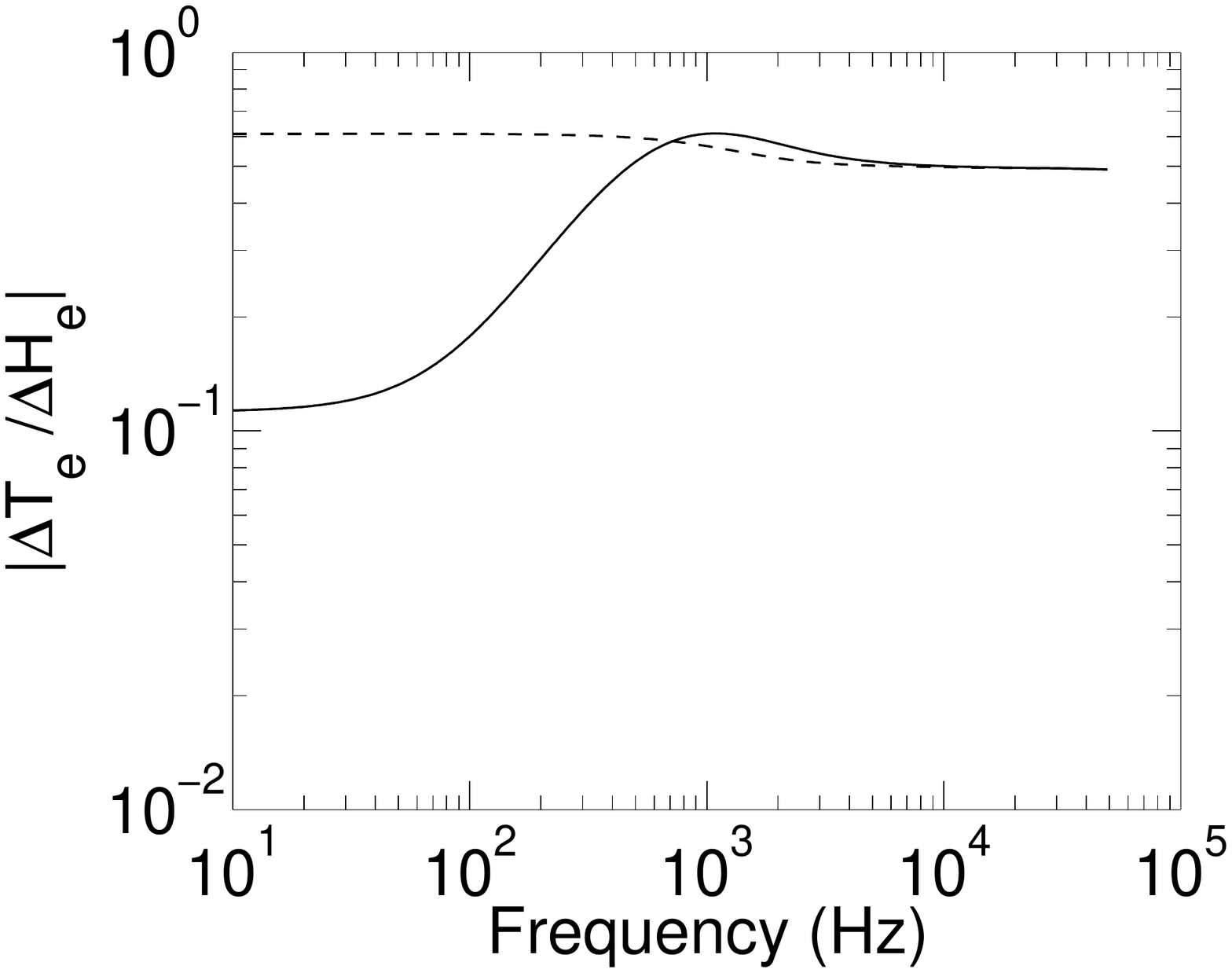}&\hspace{-0.6cm}
\includegraphics[width=0.265\textwidth]{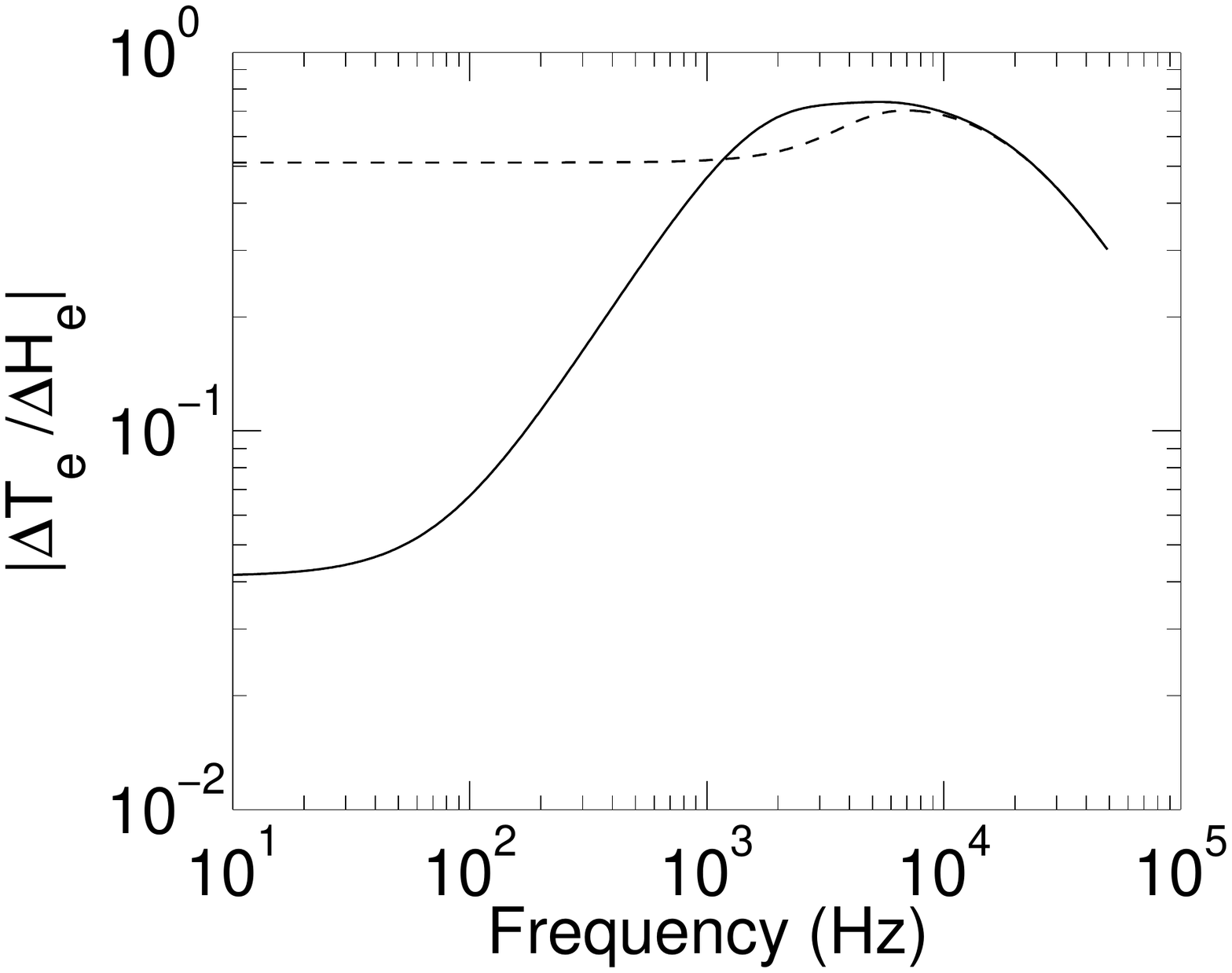}\\\vspace{-2.8cm}\hspace{-0.35cm}
\includegraphics[width=0.265\textwidth]{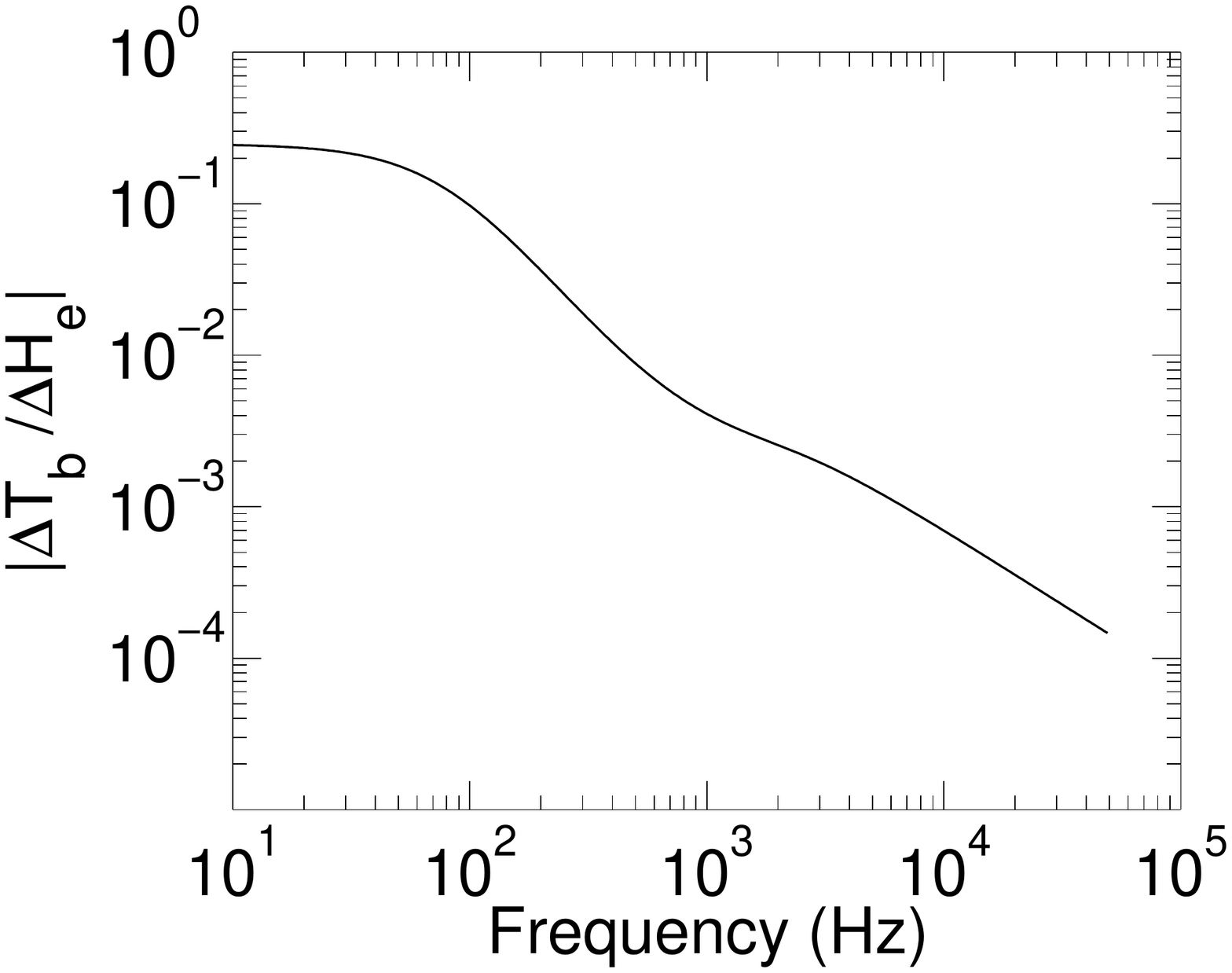}&\hspace{-0.6cm}
\includegraphics[width=0.265\textwidth]{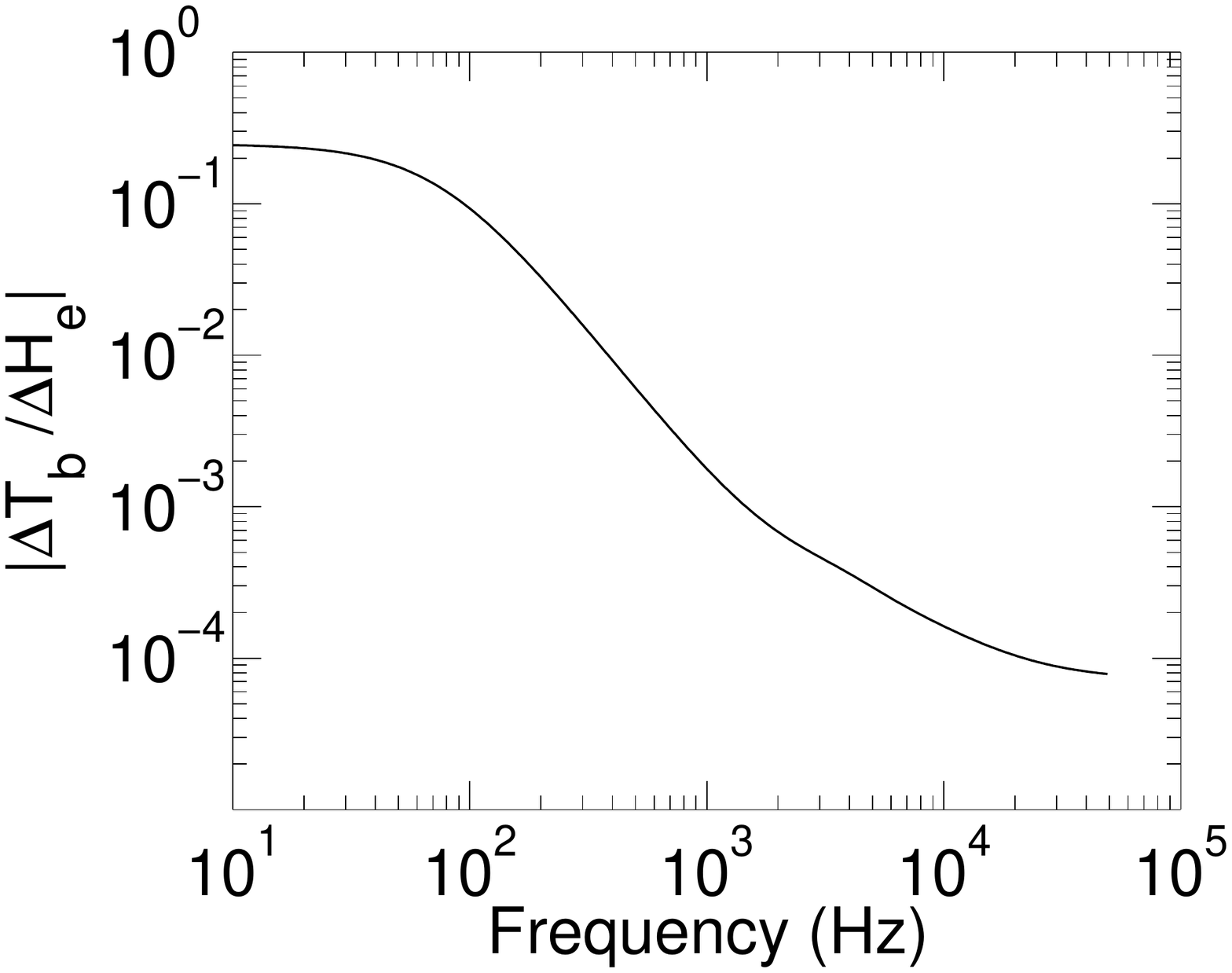}\\\vspace{-2.8cm}\hspace{-0.35cm}
\includegraphics[width=0.265\textwidth]{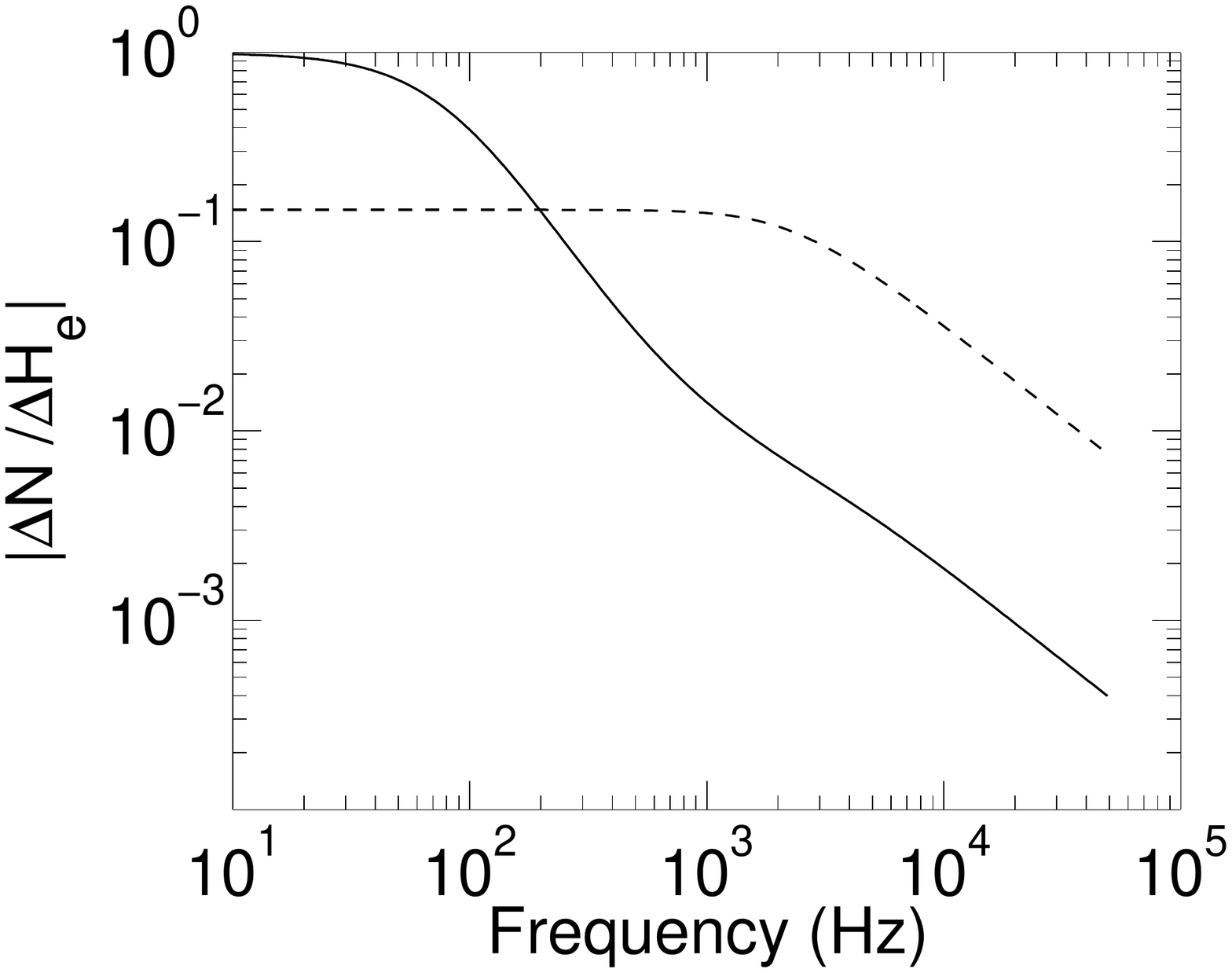}&\hspace{-0.6cm}
\includegraphics[width=0.265\textwidth]{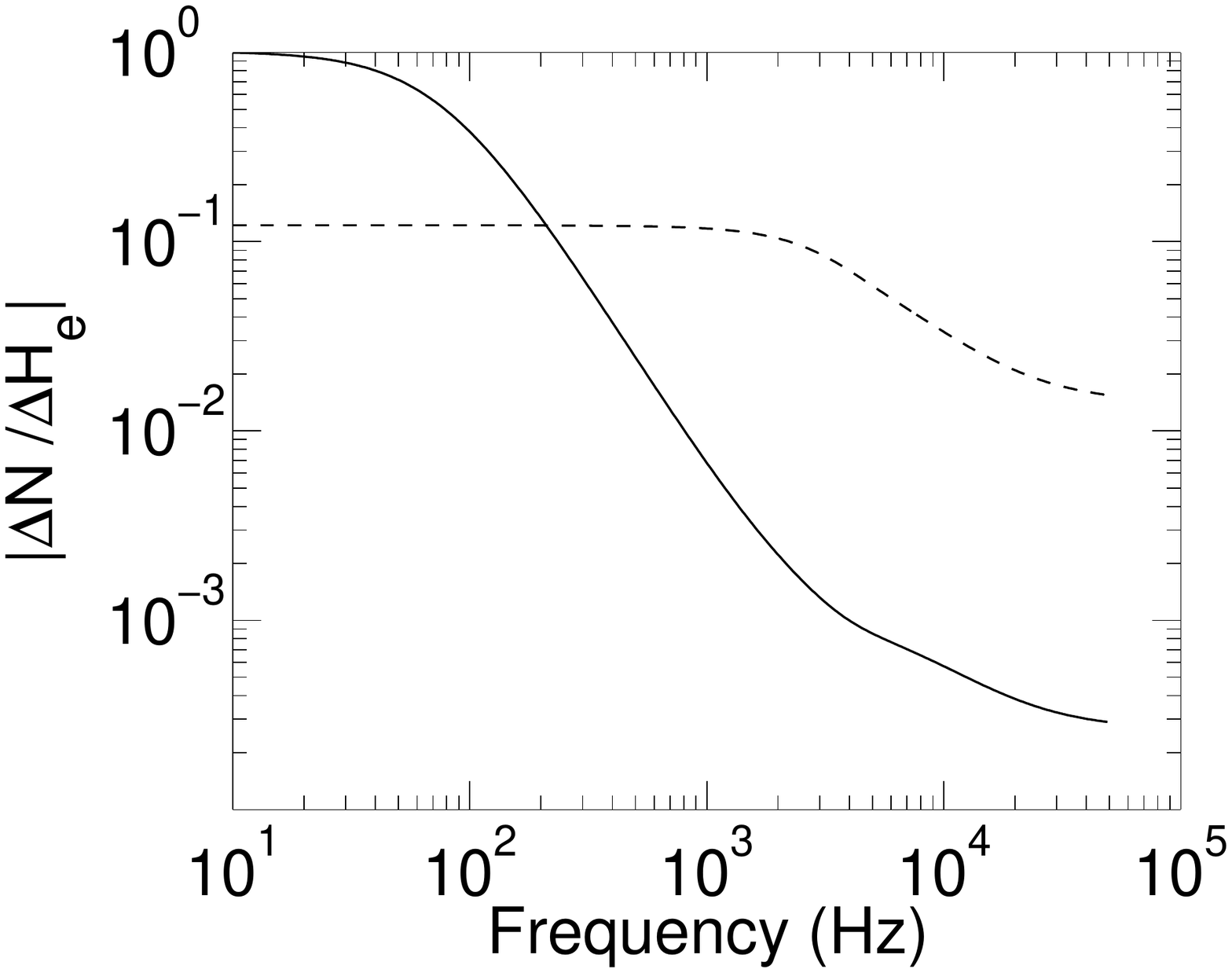}\\\vspace{0.80cm}\hspace{-0.35cm}
\end{array}$
\end{center}
\caption{Variation of $| \Delta T_e/ \Delta H_e |$, $| 
\Delta T_b / \Delta H_e |$  and the energy integrated normalised
photon density variation $ | \Delta N/ \Delta H_e | $ as
a function of frequency. Here $\Delta N$ is the normalised 
integrated photon density variation
in the energy band $2$ - $60$ keV
i.e. $\Delta N = \int n_\gamma \Delta n_\gamma dE/ \int n_\gamma dE$. 
Left column is for
the ``soft  spectrum (Model B) while the right column is for the ``
hard  spectrum (Model A).
The dotted lines are for $\eta = 0$, while the solid lines are for 
$\eta = \eta_{max} = 0.59$ (soft spectrum)
and $0.28$  (hard spectrum).}
\label{cortemp_freq}
\end{figure}

\begin{figure}\vspace{-1.7cm}
\begin{center}$
\begin{array}{cc}\vspace{-2.8cm}\hspace{-0.35cm}
\includegraphics[width=0.265\textwidth]{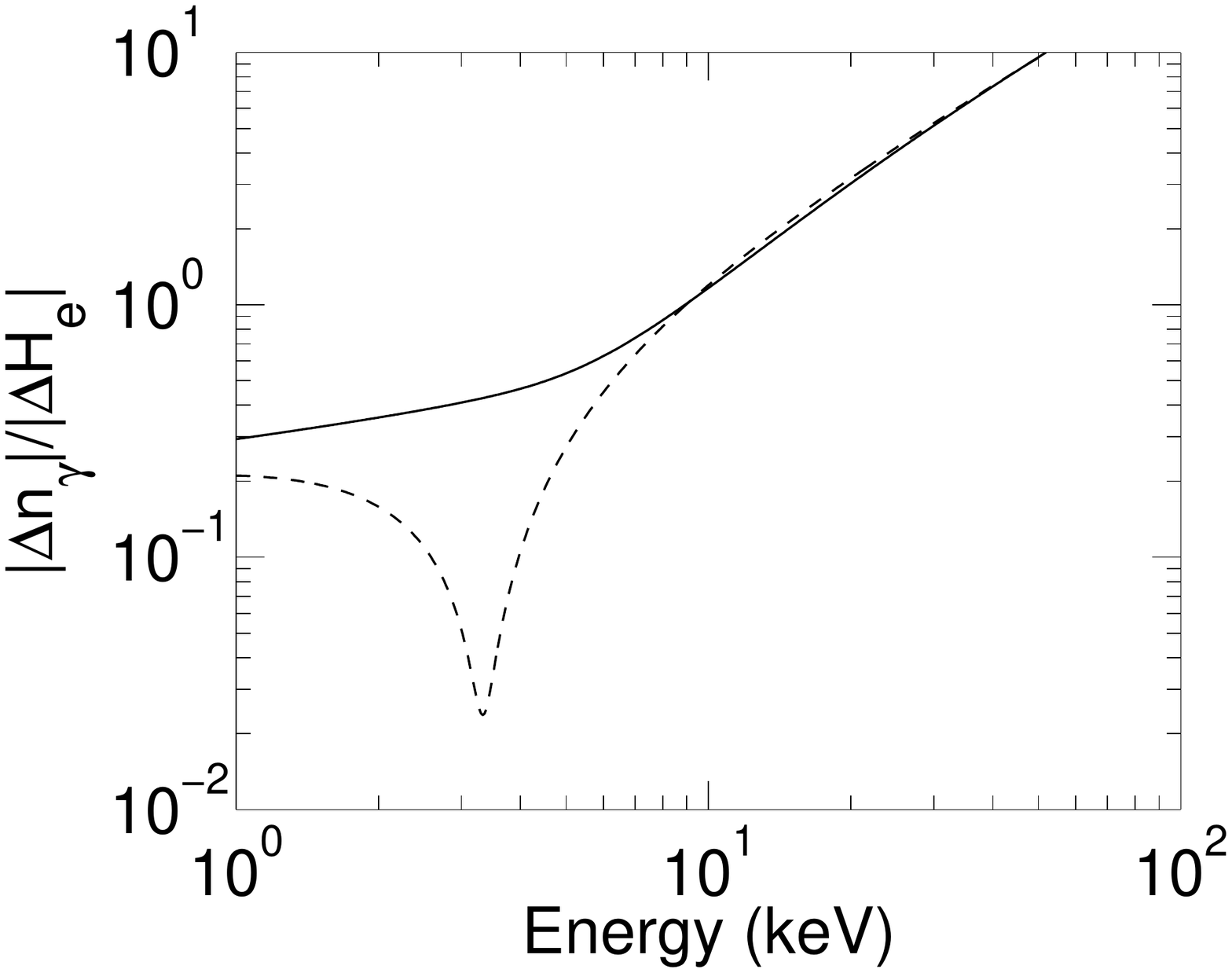}&\hspace{-0.6cm}
\includegraphics[width=0.265\textwidth]{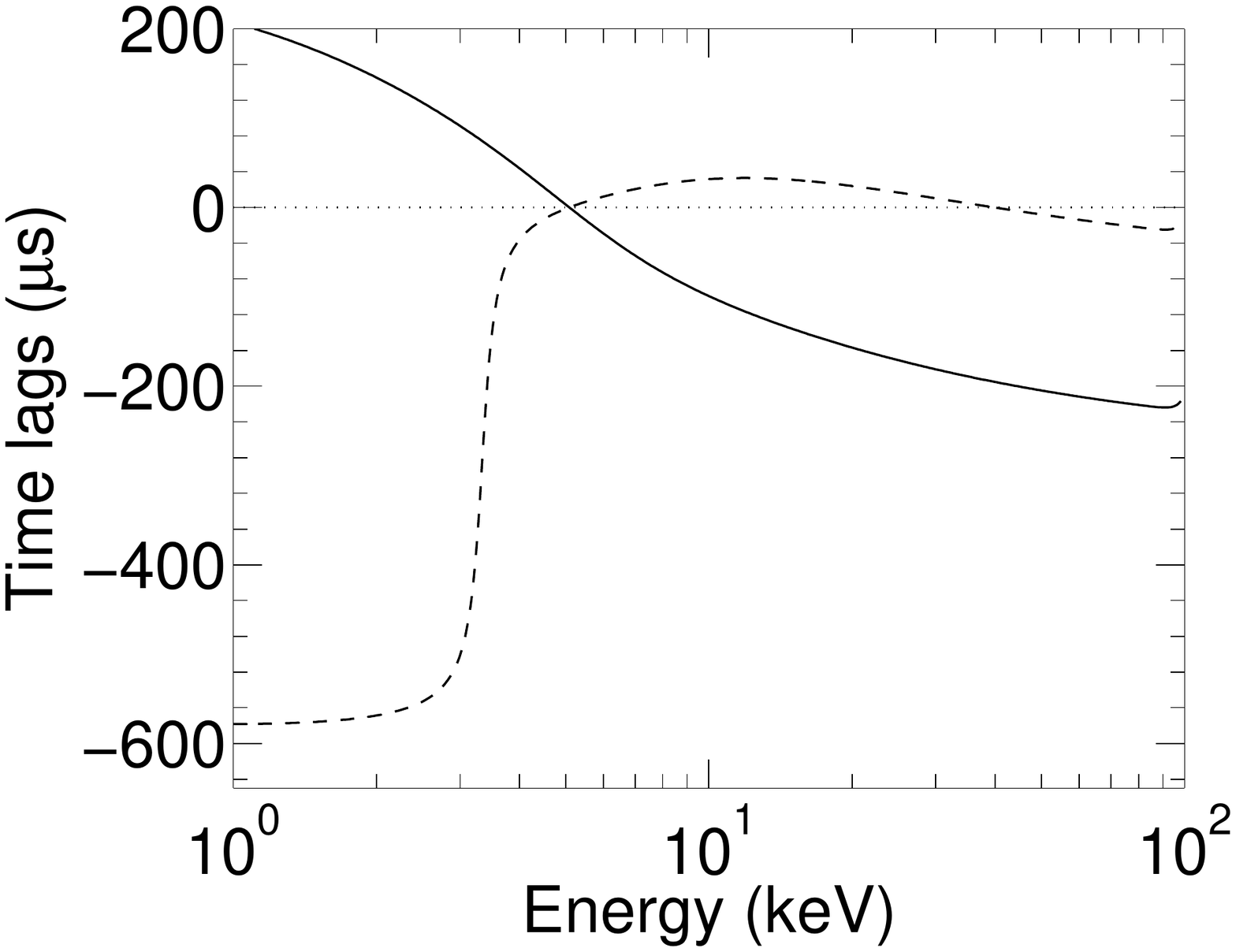}\\\vspace{-1.6cm}\hspace{-0.35cm}
\includegraphics[width=0.265\textwidth]{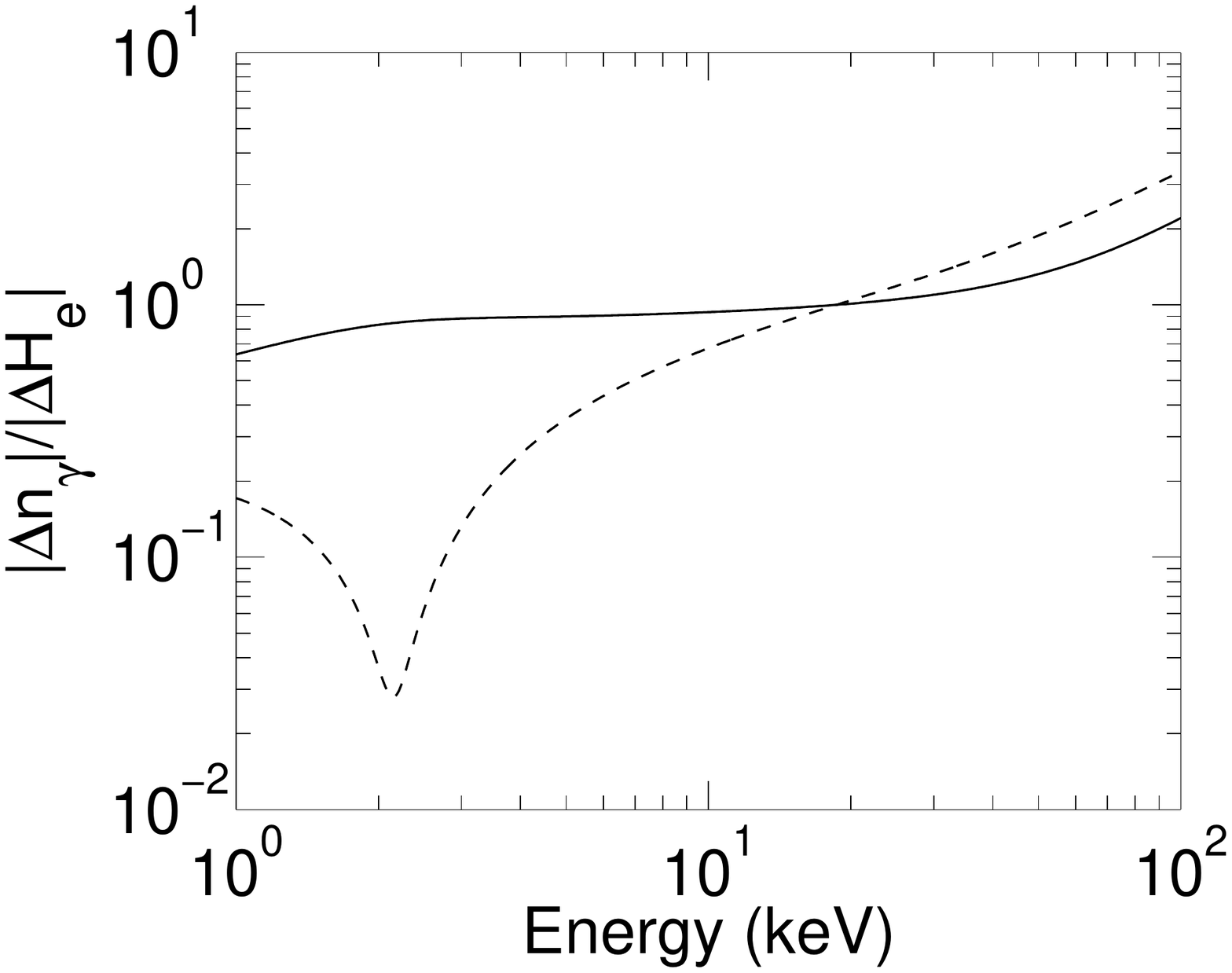}&\hspace{-0.6cm}
\includegraphics[width=0.265\textwidth]{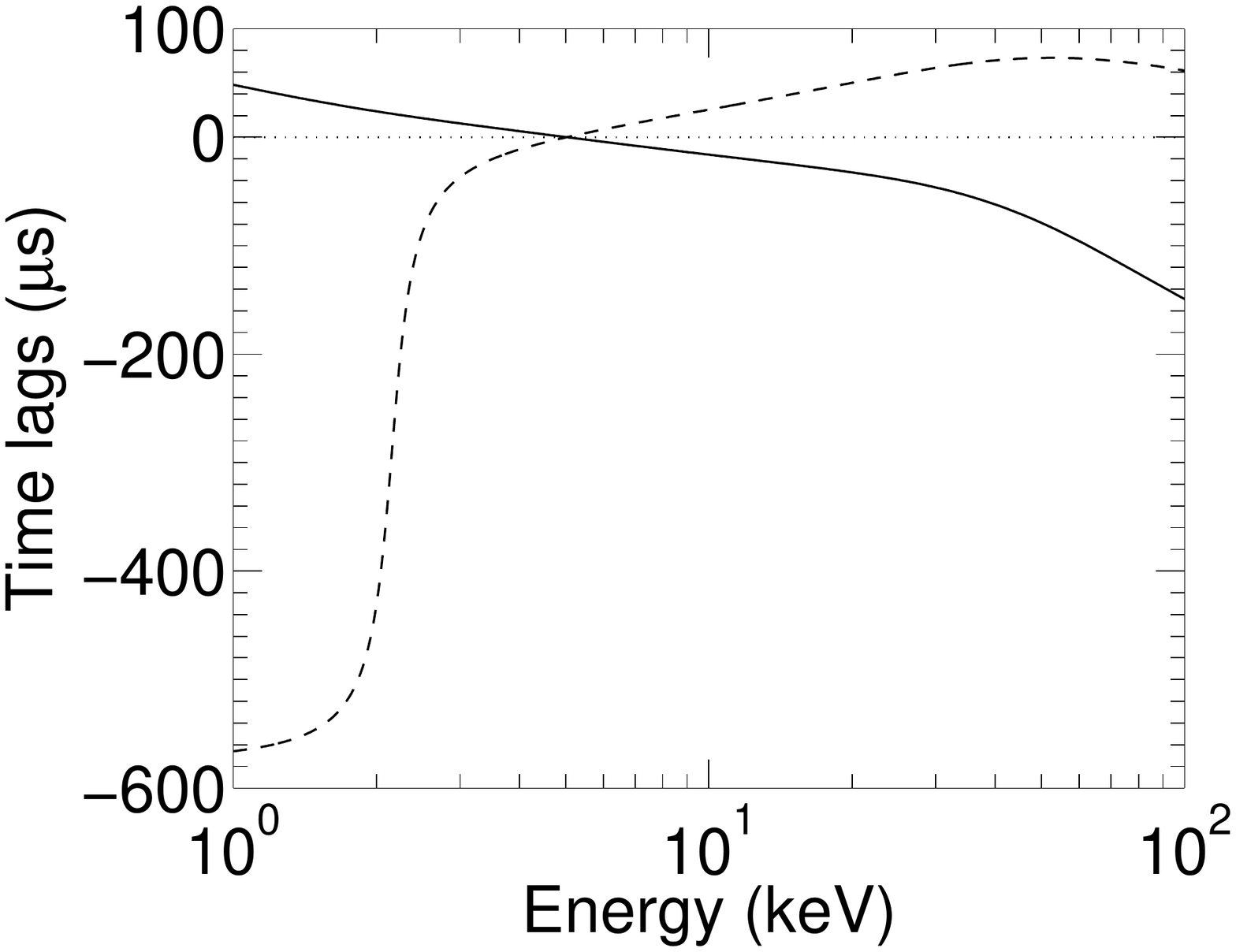}\\
\end{array}$
\end{center}
\caption{Variation of $| \Delta n_\gamma|/ |\Delta H_e |$ and 
time-lag as a function of
photon energy.  The top two figures are
for the ``soft  spectrum (Model B) while the bottom two are  
for the ``hard  spectrum (Model A). The dotted lines are for $\eta = 0$, 
while the solid lines are for 
$\eta = \eta_{max} = 0.59$ (soft spectrum)
and $0.28$  (hard spectrum).
For $\eta = 0$, there is a  pivot point at low energies $E \sim 3$ keV, 
while for $\eta = \eta_{max}$ there is no pivot point.
For $\eta = 0$  the time lags increase with energy (i.e. hard lag) 
while for $\eta = \eta_{max}$ the trend reverses and
the time lags decrease with energy (i.e. soft lag).
}
\label{cortemp_Energy}
\end{figure}

\subsection{Variation of the seed photon temperature}

Before studying the dependence with energy, we consider the behaviour of the
system for different oscillation frequencies. The variation of the
seed photon temperature leads to a variation of the corona temperature
and the ratio $| \Delta T_e/ \Delta T_b  |$ has been plotted in
the top panel of Figure \ref{seedtemp_freq} for the ``hard  and ``soft  
spectra. At low frequencies the coronal temperature responds to the
seed photon variation with $| \Delta T_e \sim 2 \Delta T_b  |$, however
at higher frequencies, the coronal temperature is unable to react to the 
rapid variation and hence varies significantly less. This is because at high
frequencies the average 
photon density inside the corona fails to respond to rapid variation in the
seed photons. This is illustrated in the right panel of Figure 
\ref{seedtemp_freq},
where the energy integrated normalised
photon density variation $ | \Delta N/ \Delta T_b | $ is plotted as
a function of frequency. $\Delta N$ is the normalised integrated photon 
density variation
in the energy band $2$ - $60$ keV, i.e. $\Delta N = \int n_\gamma 
\Delta n_\gamma dE/ \int n_\gamma dE$.
For both the ``soft''  and ``hard''  spectra, the photon density variation 
decreases sharply for
frequencies larger than $10^3$ Hz. 
The photon density variation in the corona will lead
to variations in the escaping photons and hence $| \Delta N |$ is proportional 
to the fractional r.m.s observed from a 
source. Thus, for the system under consideration, fluctuations 
with r.m.s $\sim 20$\% is unlikely
for frequencies $> 10^4$ Hz, since that would imply large values of 
$\Delta T_b  \sim 1$.
For these curves, the size of the system is $L = 5$ kms, and changing 
$L$ will lead to a proportional 
horizontal shift in frequency. For $L = 250$ kms  $| \Delta N 
/\Delta T_b | \sim 0.2$ 
for a frequency of $\sim 10^3$ Hz. This implies that the observed kHz QPO, 
restricts the size to be less than    $L < 250$ kms, which is similar 
to the constrain obtained
by light travel time arguments of $L < c \delta T \sim 300$ kms.

The energy dependence of the photon density variation 
$\Delta n_\gamma/\Delta T_b $ is
shown in Figure \ref{seedtemp_Energy} for the ``hard''  and ``soft''  
spectra. $\Delta n_\gamma$ is
equal to the normalised r.m.s that can be obtained from observations. 
The frequency of
oscillation is $850$ Hz. There is a pivot point where
the photon variation goes to zero and it is at high energies 
$E \sim 10$-$20$ keV. 
Below
the pivot point, the variation is nearly constant or 
decreasing with energy.  Also
shown in Figure \ref{seedtemp_Energy} is the expected 
lag between different energies.
The time lag is an increasing function of energy 
while implies that the hard photon
lag the soft ones (i.e. a  hard lag).

\subsection{Variation of the coronal heating rate}

\begin{figure*}
\begin{center}$
\begin{array}{ccc}
\includegraphics[width=0.356\textwidth]{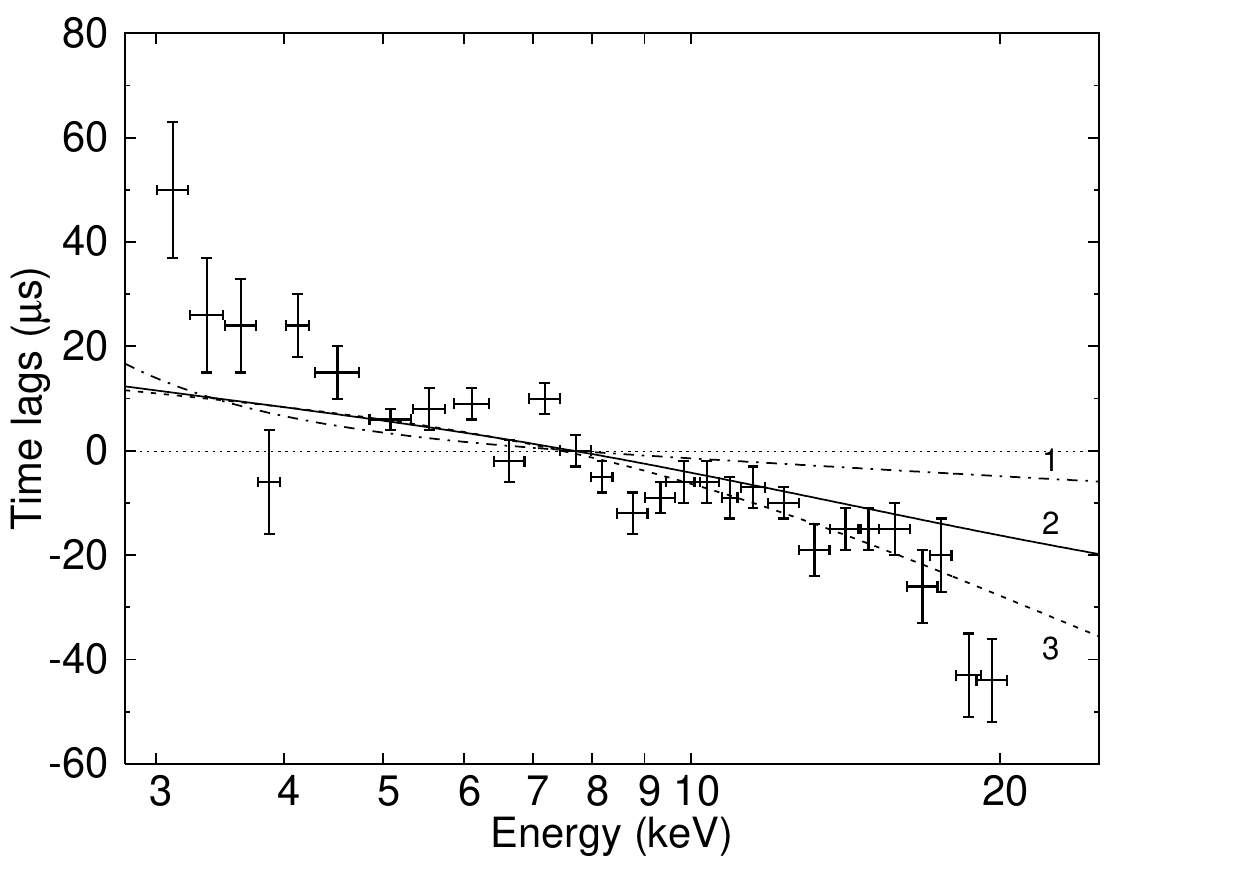}&\hspace{-0.6cm}
\includegraphics[width=0.356\textwidth]{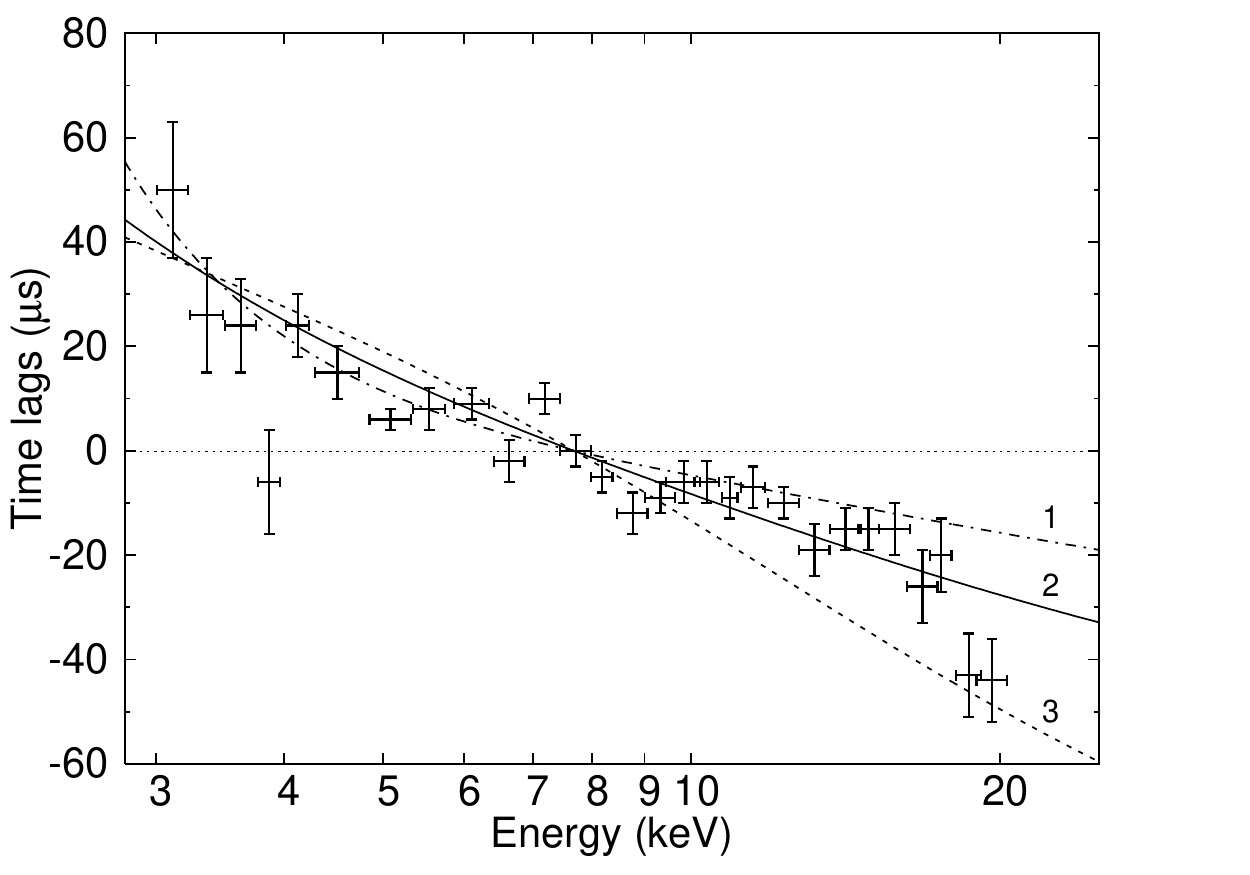}&\hspace{-0.6cm}
\includegraphics[width=0.356\textwidth]{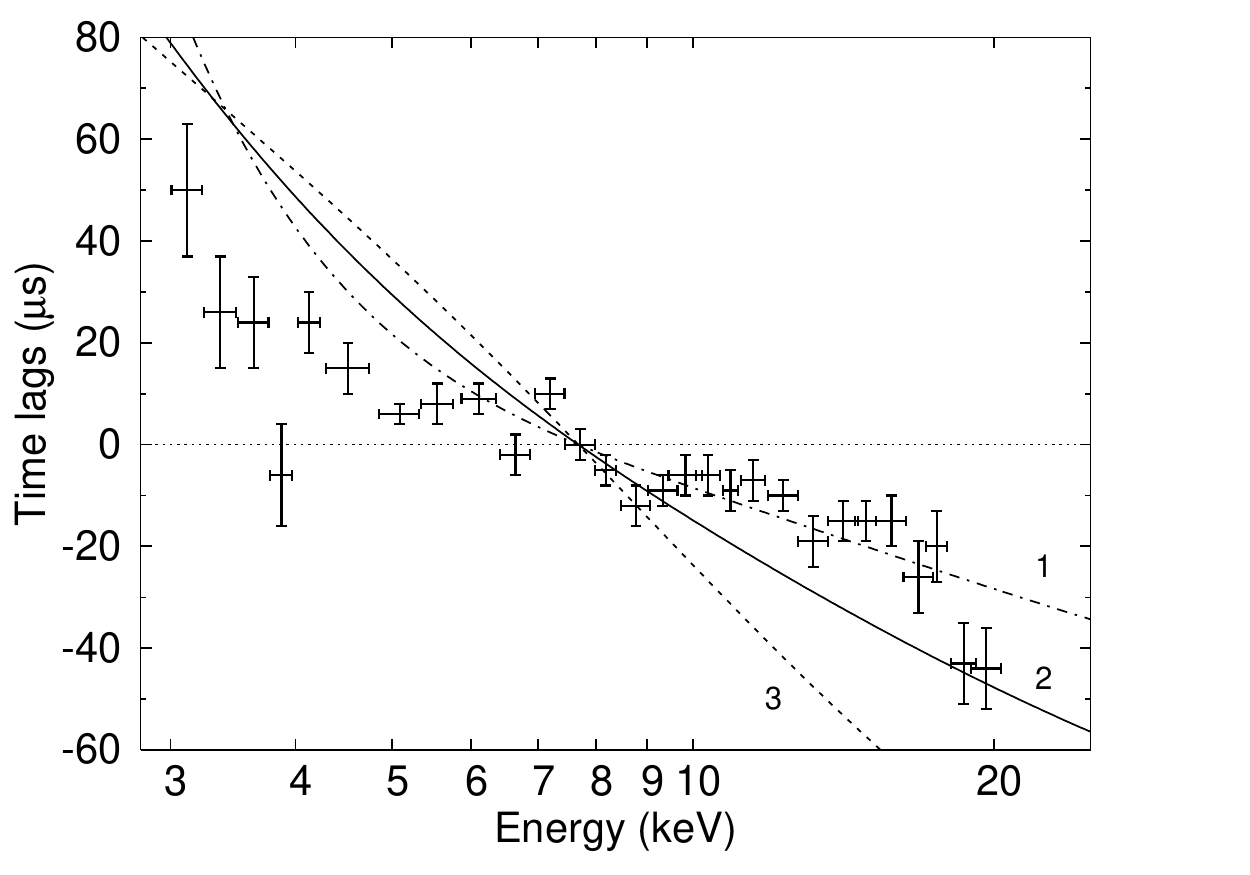}\\
\end{array}$
\end{center}
\caption{Time lags versus energy for the March 3rd observation of 4U 1608-52.
The data is the same for all three panels.
 The left panel is for $L=0.3$ km and the curves marked 1,2, and 3 are
for $\eta$ 0.3, 0.5, and 0.57 respectively. 
The middle and right panels are for $L=1.0$ and $2.0$ km and the 
curves marked 1,2, and 3 are 0.3
, 0.4, and 0.5 respectively.  The best description of the data is obtained
for $L = 1$ km and $\eta = 0.4$ (solid line in the middle panel). 
}
\label{lag_1608}
\end{figure*}

\begin{figure*}
\begin{center}$
\begin{array}{ccc}
\includegraphics[width=0.356\textwidth]{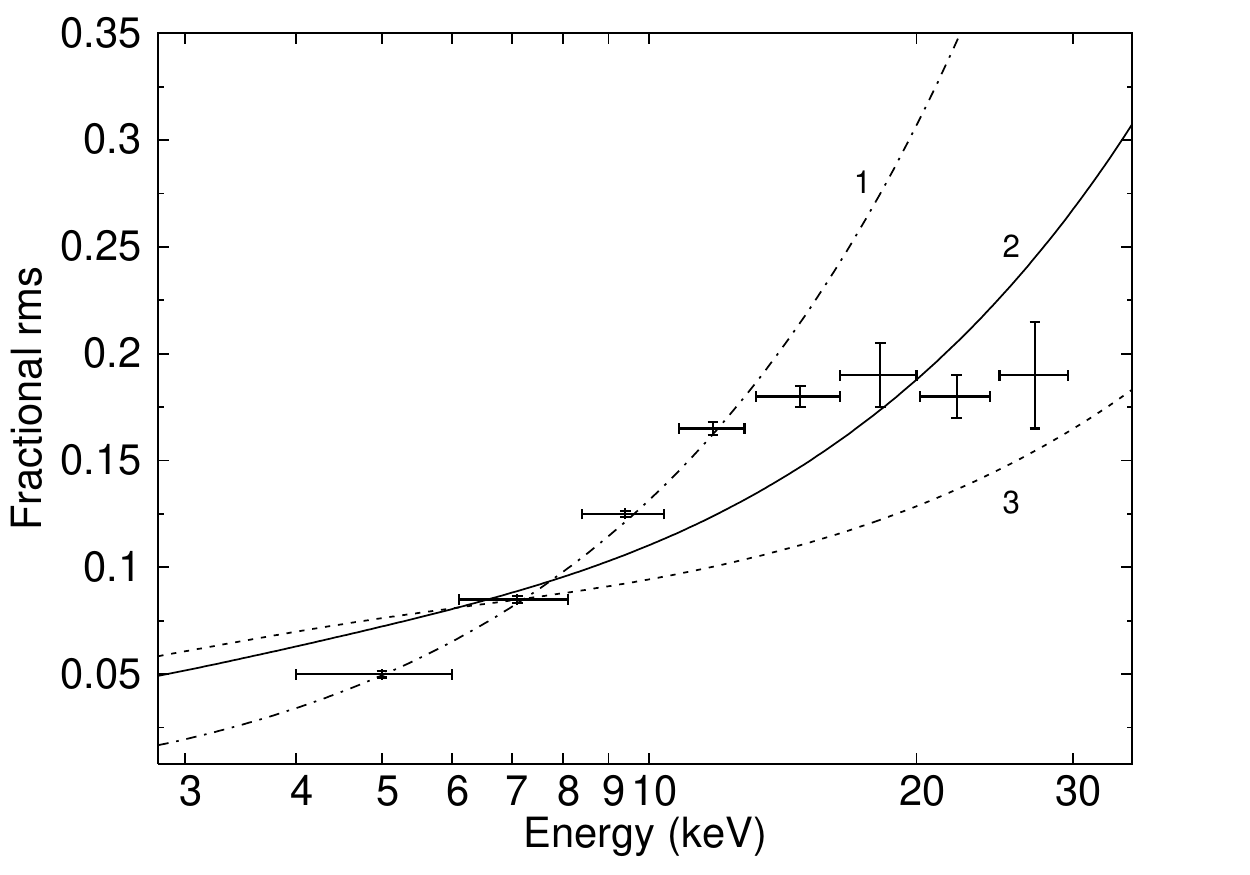}&\hspace{-0.6cm}
\includegraphics[width=0.356\textwidth]{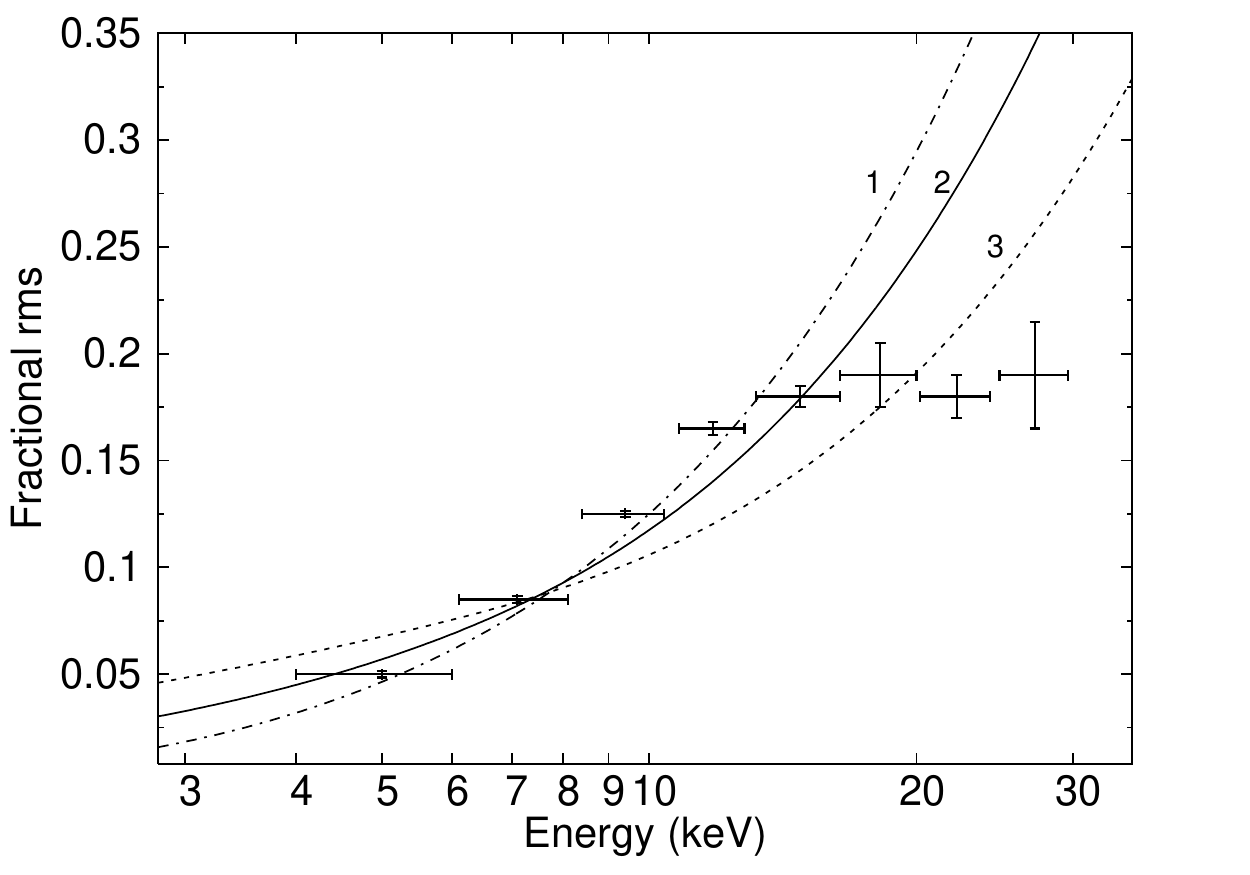}&\hspace{-0.6cm}
\includegraphics[width=0.356\textwidth]{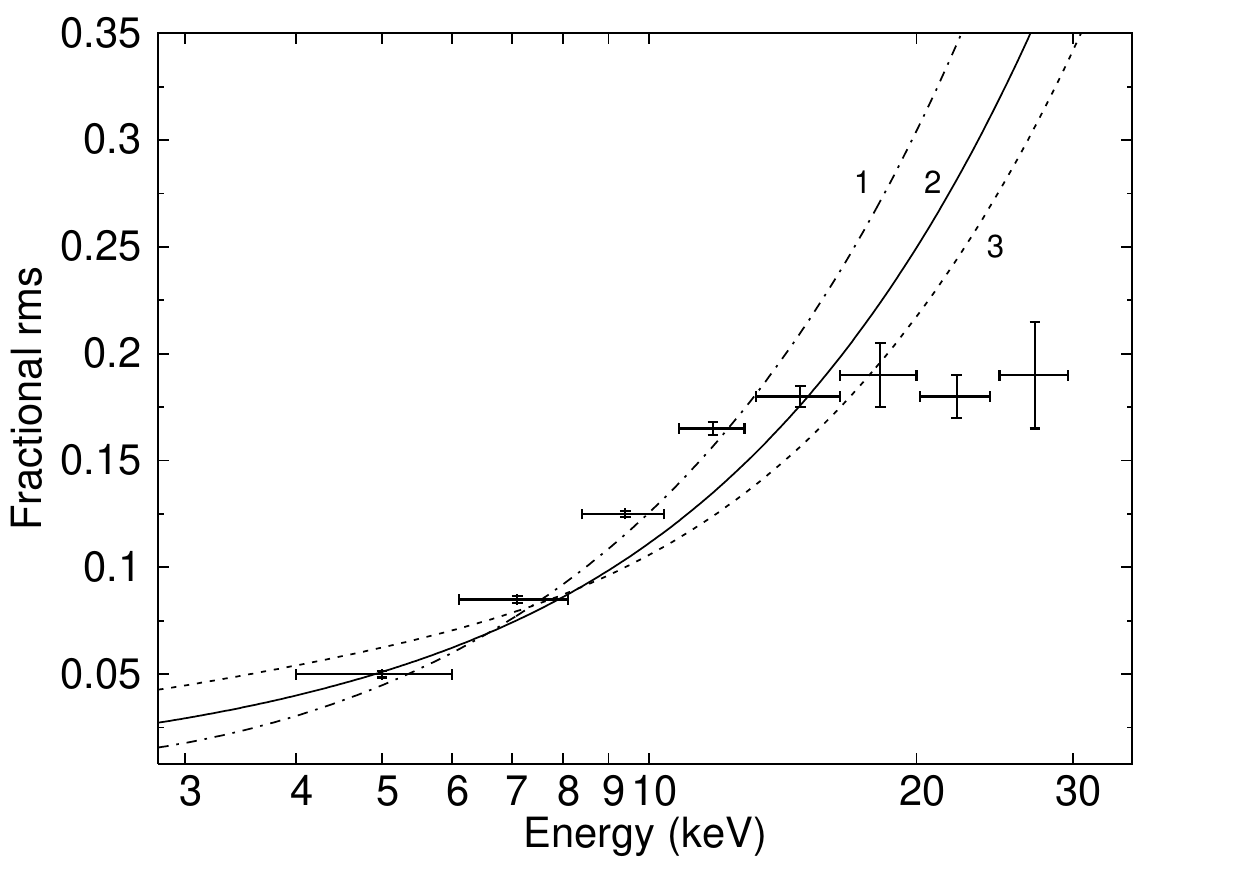}\\
\end{array}$
\end{center}
\caption{Fractional rms versus energy for the March 3rd observation of 4U 1608-52.
The predicted lines correspond to the same parameters as used for the lines
in Figure \ref{lag_1608}. Note that for energies $< 20$ keV, the parameters
that describe the lag variation ($L = 1$ km and $\eta = 0.4$) also describes
the r.m.s versus energy (solid line in the middle panel).
}
\label{rms_1608}
\end{figure*}

Variation of the coronal heating rate, $\Delta H_e$ will induce a 
variation of the coronal temperature, $\Delta T_e$ and
their ratio as a function of frequency  is plotted in Figure  
\ref{cortemp_freq}. If there is a fraction of photons which impinge back into
the soft photon source (i.e. $\eta \neq 0$), then there will be a 
variation of the seed photon temperature as well and this is
plotted in the middle panel of Figure  \ref{cortemp_freq}. The bottom panel 
shows the energy integrated 
photon density variation $ | \Delta N/ \Delta H_e | $ 
as a function of frequency.
 For the Figure, we have considered two extreme cases of $\eta = 0$ and  
$\eta = \eta_{max}$, where $\eta_{max}$ is the maximum allowed fraction. 

There is not much qualitative differences  between the results of the 
``soft''  and ``hard''  state spectra (left and right
columns of Figure \ref{cortemp_freq}) which implies that the 
variability nature of such systems is not sensitive to
the time averaged spectrum.  However, there are important differences 
in the predicted variability of a system with
$\eta = 0$ and $\eta = \eta_{max}$.  For $\eta = 0$ the energy integrated 
photon variability, $|\Delta N/ \Delta H_e|$ is a constant and
decreases for frequencies higher than $10^3$ Hz, fairly similar to the 
results obtained for the case when the
seed photon temperature is varied (Figure \ref{seedtemp_freq}). 
For $\eta=\eta_{max}$, 
$|\Delta N/ \Delta H_e|$ starts decreasing for frequencies 
$> 100$ Hz and is $\sim 0.05$ at $\sim 1$ kHz. For any observed
kHz QPO with fractional r.m.s $> 5$\%, would require an unphysically 
large variation in the heating rate
$|\Delta H_e| > 1$. For such a system to be viable its size 
should be much smaller than $L = 5$ kms assumed here.
Note that this is a more stringent limit that that obtained 
from light travel time arguments of $L < c \delta T \sim 300$ kms.

The variation of $| \Delta n_\gamma/ \Delta H_e |$ 
and time-lag as a function of
photon energy are shown in Figure \ref{cortemp_Energy}.
Again, there are qualitative differences between
the case when $\eta = 0$ and $\eta= \eta_{max}$. 
For $\eta = 0$, the pivot point is at low
energies while no pivoting is seen for $\eta= \eta_{max}$. 
More strikingly, the time lag behaviour
changes. While for $\eta = 0$, the hard photons lag the 
soft ones, the reverse occurs for
$\eta= \eta_{max}$. 
Thus, in this interpretation soft lags occur only when a significant fraction
of the Comptonized photons impinge back into the input photon source.

\subsection{Comparison with the energy dependent time lags of 4U 1608-52}

As discovered by \citet{Vaughan-etal1997}, the RXTE observation of the
$\sim 850$ Hz QPO of 4U 1608-52 on
3rd March 1996, remains one of the best cases of soft lag  detection in
kHz QPOs. Recently,  \cite{Barret2013} have estimated the time
lag as a function of energy for more than 20 energy bins and have
interpreted some of the qualitative features as due to reverberation lags.
Thus, we choose this observation set as an example to illustrate how
these time-lags can be interpreted in terms of Comptonization time delays
and the possible constraints that can be obtained.  

As shown in the previous section, soft time lags are expected only when
the variation is in the coronal heating rate and with a significant fraction of 
photons impinging back into the input photon source.  
A detailed analysis can be undertaken only when
the different components of the time-averaged spectrum are well constrained.
In particular, the spectra of these sources require apart from a primary
thermal component, a soft component modelled as a disk black body and
a broad Iron line, which implies the presence of a reflected component.
The $3$-$10$ keV low resolution spectrum obtained from the PCA observations 
are not sufficient to constrain these complex parameters. 
The absorption column density is undetermined and is often fixed to the
Galactic value which might be an underestimate. Nevertheless, the primary
component is the thermal Comptonization and when we fit a single thermal
component model to the time averaged spectrum of this observation, we find
that the model fits the data with  a systematic uncertainty of 10\%. The
parameters obtained from such a fit , electron temperature $kT_e = 2.6$ keV, 
optical depth  $\tau^2$ = 40 and soft photon temperature of $0.7$ keV are
similar to those obtained from  more sophisticated analysis 
\cite[e.g.][]{Barret2013}. We defer a detailed analysis taking into
account different spectral parameters to a latter work. Apart from 
the spectral parameters, there are two other important parameters which 
determine the time-lag versus energy. These are the size of the Comptonizing
region, $L$  and the fraction, $\eta$ of photons impinging back into the input 
source. These two parameters can be constrained by comparison with the
observed values.

The three panel in Figure \ref{lag_1608} all show the observed time-lag
versus energy for the observation which are compared with model predictions. 
The curves in the left, middle and right panels correspond to 
three different sizes of the Comptonizing regions, 
$L = 0.3$, $1.0$ and  $2.0$ kms respectively. The curves marked 1, 2, and 3 
are for three different values of the fraction $\eta$ mentioned in the 
caption. The best representation of the data occurs when $L = 1.0$ km 
and $\eta = 0.4$ (solid line in the middle panel). For smaller values of
$L = 0.3$ kms, the magnitude of the time-lag is smaller than observed,
while for larger values, $L = 2$ kms it is larger. Thus we demonstrate that
time-lag analysis can well constrain the size of the Comptonizing region.

The model also predicts the r.m.s. versus energy for the QPO. In such a linear
model, the normalisation of the r.m.s. scales with the magnitude of the
perturbation $\Delta T_e$, but the shape of the curve is determined by the
spectral parameters, $\eta$ and the size $L$ of the system. 
Figure \ref{rms_1608} shows the variation of r.m.s versus energy for the
same observation.  
The data points have been taken from \cite{Berger-etal1996}. The lines in 
the three panels are the predicted r.m.s for the model parameters used in
Figure \ref{lag_1608}. It is interesting to note that for photon energies
$< 20$ keV, the predicted variation matches well with the observed values
for  $L = 1$ km and $\eta = 0.4$ (solid line of second panel in Figure
\ref{rms_1608}). The observed r.m.s. is lower for higher energies possibly
due to the presence of an additional constant hard component.

\section {Summary and Discussion}

The linearised time dependent Kompaneets equation describing thermal
Comptonization has been solved self consistently taking into account energy
balance of the plasma and soft photon source. When the perturbation is due
to temperature variation in the soft photon source, the time-lag between 
photons of different energies is found to be hard. The fractional r.m.s 
as a function of energy reveals a pivot point at high energies and is nearly
constant at lower energies. When the perturbation is in the heating rate of
the plasma the pivot point is located at low energies. There is a major
qualitative change in the behaviour of the system if a significant fraction of
the photons impinge back into the soft photon source. In this case, the
time-lag is found to be soft and the r.m.s increases monotonically with energy.
Such behaviour has been reported for the kHz oscillations in LMXBs. As an
example, we compare
the model predictions with the time-lag and r.m.s observed for the 850 Hz 
oscillation of 4U 1608-52 and show that the size of the Comptonizing region 
can be constrained to $\sim 1$ km. 

The analysis shows that modelling the time-lag and r.m.s variation as
a function of energy for kHz QPOs is a powerful technique which can in
principle constrain the size and geometry of the source. A more detailed
analysis will require a careful modelling of the time-dependent spectra. 
Modelling the time-lag as well as the r.m.s at different QPO frequency and
taking into account the corresponding spectral changes, may provide information
regarding how the size and geometry of the system changes. Moreover, while
the time-lags for the upper kHz QPO have not been constrained, they are known
to be significantly different than those of the lower kHz QPO and hence should
provide additional constraints. We defer such a 
detailed analysis to explain 
the observations reported by \cite{deAvellar-etal2013} to a latter work. 
One may also need to take into account additional effects such as 
reverberation lags. The Iron line
component and the soft disk emission may oscillate in response to the plasma
variation after a time-lag. 
However, it is important to note that even for
relatively small plasma size of $\sim 1 $km, the time-lags due to Comptonization
is of the order of $\sim 50$  microseconds and hence they are expected to
dominate. The more subtle and complicated General Relativistic effects may
also be needed to be incorporated in order to correctly model such systems.

While the qualitative results such as the presence of soft lags do not
depend on the time averaged spectral parameters, the constraints obtained on the
size and geometry do depend on their precise values. Thus, it is important
to obtain broad band spectral data during the time when a kHz QPO has been
observed. This will be possible by the upcoming multi-wavelength satellite
ASTROSAT\footnote{http://astrosat.iucaa.in} \citep{Agrawal2006} 
where the Large array X-ray proportional counter will detect
the QPO while it and the other X-ray instruments will provide
broad band spectra.

\label{lastpage}

\end{document}